\documentclass[a4paper]{amsart}

\usepackage[leqno]{amsmath}
\usepackage{amssymb,amsthm}
\usepackage{mathtools}
\usepackage{bm}
\usepackage[a4paper,margin=28mm]{geometry}
\usepackage{microtype}
\usepackage[utf8]{inputenc}
\usepackage[T1]{fontenc}
\usepackage{lmodern}

\usepackage[english]{babel}

\usepackage[mathcal]{euscript}
\usepackage{mathrsfs}

\allowdisplaybreaks[1]
\usepackage[pdfauthor={Huang and Luan and Xue}, pdftitle={Super Howe duality and rigidity of hybrid Fourier transforms},pdfcreator={pdflatex}]{hyperref}

\hypersetup{hidelinks}
\usepackage{cite,doi,url}
 \usepackage{arydshln}
\usepackage{verbatim,enumerate}
\usepackage{graphicx,picinpar,pdfpages}
\usepackage{xcolor}

\usepackage{tikz, tikz-3dplot, pgfplots}
\usetikzlibrary[positioning,patterns]
\pgfplotsset{compat=1.8}
\usetikzlibrary{arrows.meta}
\usetikzlibrary{matrix}
\usetikzlibrary{babel}
\usepackage{tikz-cd}
\usetikzlibrary{arrows}

\theoremstyle{plain}
\newtheorem{thm}{Theorem}[section]
\newtheorem{lemma}[thm]{Lemma}
\newtheorem{prop}[thm]{Proposition}

\newtheorem{corl}[thm]{Corollary}

\theoremstyle{definition}
\newtheorem{defn}[thm]{Definition}
\newtheorem{exmp}[thm]{Example}

\theoremstyle{remark}
\newtheorem{remark}[thm]{Remark}

\numberwithin{equation}{section}

\newcommand{\compi}{{i\mkern1mu}} 
\newcommand{\mdot}{\,\text{.}}
\newcommand{\mcomma}{\,\text{,}}

\newcommand{\bbC}{\mathbb C}
\newcommand{\bbE}{\mathbb E}

\newcommand{\bbN}{\mathbb N}

\newcommand{\bbR}{\mathbb R}

\newcommand{\caF}{\mathcal F}

\newcommand{\caH}{\mathcal H}
\newcommand{\caP}{\mathcal P}
\newcommand{\caS}{\mathcal S}
\newcommand{\caT}{\mathcal T}

\newcommand{\frg}{\mathfrak g} 
\newcommand{\fh}{\mathfrak h}

\newcommand{\fl}{\mathfrak l}

\newcommand{\fs}{\mathfrak s}

\newcommand{\fgl}{\mathfrak{gl}}
\newcommand{\fosp}{\mathfrak{osp}}
\newcommand{\fsl}{\mathfrak{sl}}
\newcommand{\fsu}{\mathfrak{su}}
\newcommand{\fso}{\mathfrak{so}}
\newcommand{\fsp}{\mathfrak{sp}}

\newcommand{\bmx}{\bm{x}}
\newcommand{\bmy}{\bm{y}}
\newcommand{\bmgrx}{\grave{\bm{x}}}
\newcommand{\bmgry}{\grave{\bm{y}}}
\newcommand{\grx}{\grave{x}}
\newcommand{\gry}{\grave{y}}

\newcommand{\stackarrows}[2]{%
  \substack{ \xrightarrow[{}]{#2}\\  \xleftarrow[#1]{} }%
}

\DeclareMathOperator{\End}{End}

\DeclareMathOperator{\Id}{Id}

\DeclareMathOperator{\vectorspan}{span}

\newcommand{\spanC}{\operatorname{span}_{\mathbb{C}}}

\newcommand{\Ub}{U_{\mathrm b}}
\newcommand{\Gf}{G_{\mathrm f}}
\newcommand{\Uf}{U_{\mathrm f}}
\newcommand{\Fhyb}{F_{m|2n}}
\newcommand{\Uhyb}{U_{m|2n}}
\newcommand{\Dn}{D_n}
\newcommand{\Pf}{P_f}
\newcommand{\Pb}{P_b}
\newcommand{\Pbf}{P_{bf}}

\begin{document}
\title{Super Howe duality and rigidity of hybrid Fourier transforms}

\author[J.-S. Huang]{Jing-Song Huang}
\address[J.-S. Huang]{Mathematics Research Center, School of Science and Engineering, The Chinese University of Hong Kong, Shenzhen, 518172, China}
\email{huangjingsong@cuhk.edu.cn}

\author[Y. Luan]{Yongzhi Luan}
\address[Y. Luan]{School of Mathematics, Shandong University, Jinan, 250100, China}
\email{luanyongzhi@email.sdu.edu.cn}

\author[Y. Xue]{Yinuo Xue}
\address[Y. Xue]{School of Science and Engineering, The Chinese University of Hong Kong, Shenzhen, 518172, China}
\email{yinuoxue@link.cuhk.edu.cn}

\date{\today}

\begin{abstract}
We characterize the hybrid Fourier transform on the Schwartz--Grassmann space $\caS(\bbR^m)\otimes\Lambda_{2n}$ by operator intertwining relations. Using the bosonic oscillator Howe decomposition and the fermionic Lefschetz decomposition, we describe the action of the hybrid Fourier transform on both tensor factors. We show that first-order covariance with respect to coordinate multiplication and differentiation determines the transform up to a scalar. For $m,n\ge1$, we prove that covariance under all even quadratic operators, together with any one nonzero mixed quadratic operator, determines the linear span of the hybrid Fourier transform and its inverse. This rigidity result is obtained by computing the ordinary commutants of the even and full quadratic oscillator actions of $\fosp(4n|2m)$. After a degree-scaling change of normalization, we make the hybrid Fourier transform unitary with respect to the occupation number inner product, and factor it into a bosonic oscillator rotation and (up to a scalar phase) an $SU(2)$ rotation in the fermionic Lefschetz representation. For $m=1$ and $n\ge2$, we show that full even quadratic covariance together with the parity-preserving cubic relation for $qQ_f$ determines the unitary transform up to a global phase and a fermion-parity phase. Finally we calculate the unitary conjugation formulas for Hubbard--Holstein Hamiltonian for a hybrid continuous-variable and discrete-variable quantum system.
\end{abstract}

\maketitle
\tableofcontents

\section{Introduction}
The classical Fourier transform is a fundamental unitary operator in the oscillator representation. In \cite{HZ19}, Huang and Zhou developed a Howe duality approach to study generalized Fourier transforms associated with definite and indefinite bilinear forms. In the setting of superspace, H. De Bie constructed a Fourier transform with a symplectic fermionic kernel and proved its basic intertwining and spectral properties in \cite{DB08}. In this paper, we work on the characterization and rigidity of this hybrid Fourier transform via operator covariance, and we settle down the parity freedom that remains when only quadratic intertwining relations are imposed.

Let us consider the Schwartz--Grassmann space
\[
 \caS(\bbR^{m|2n}) \coloneqq \caS(\bbR^m)\otimes_{\bbC}\Lambda_{2n} \mcomma
\]
where $\caS(\bbR^m)$ is the Schwartz space on $\bbR^m$, and $\Lambda_{2n}$ is the complex Grassmann algebra on $2n$ generators. Following the conventions fixed in Sections \ref{sec:bosonic} and \ref{sec:grassmann}, we write the hybrid Fourier transform in the factorized form
\[
 \Fhyb^\sigma = \Ub^\sigma\otimes\Gf^\sigma \mcomma
 \qquad \sigma\in\{+1,-1\} \mdot
\]
Here, $\Ub^+$ (respectively, $\Ub^-$) is the ordinary bosonic Fourier transform (respectively, its inverse), and $\Gf^\sigma$ is the fermionic Fourier transform defined by the normalized Berezin integral. We use the term \textbf{hybrid} to emphasize that the transform combines the continuous bosonic factor $\caS(\bbR^m)$ with the finite-dimensional fermionic factor $\Lambda_{2n}$. The tensor product $\Fhyb^\sigma$ coincides with De Bie's superspace Fourier transform in \cite{DB08}.

In our study, we use two different representation-theoretic structures. First, on the coordinate level, the dual pair $(\fosp(m|2n), \fsl_2)$ governs the radial covariance and harmonic analysis on superspace (cf. \cite{C13, Z08}). On the purely fermionic factor $\Lambda_{2n}$, this specializes to the finite-dimensional symplectic Lefschetz decomposition
\[
 \Lambda_{2n} \cong \bigoplus_{k=0}^n
 V_{n-k}\otimes\caH_k^f \mdot
\]
Combining this Lefschetz decomposition with the classical bosonic oscillator Howe decomposition (cf. \cite{How95, HZ19}), we obtain the explicit block decomposition and eigenvalues of $\Fhyb^\sigma$ on the irreducible summands of both factors. Note that on a mixed superspace, the usual Fischer direct sum decomposition can fail when the even superdimension $M = m - 2n$ is nonpositive, whereas the finite-dimensional fermionic Lefschetz decomposition always holds (cf. \cite{C13}). Second, to prove our quadratic rigidity theorems in Section \ref{sec:hybrid}, we work with the Lie superalgebra of quadratic operators in the Weyl--Clifford algebra, which is isomorphic to $\fosp(4n|2m)$ with even part $\fsp(2m) \oplus \fso(4n)$. In particular, our rigidity proof does not rely on the coordinate Fischer decomposition on mixed superspace.

We first study the fermionic factor $\Lambda_{2n}$ in Section \ref{sec:grassmann}. We prove that first-order covariance with respect to Grassmann multiplication and contraction determines $\Gf^+$ up to a scalar. When $n \geq 2$, covariance under pure creation and annihilation quadratics determines the two-dimensional subspace $\spanC\{\Gf^+, \Gf^-\}$, whereas for $n = 1$ the solution space is five-dimensional. Using the $(\fsl_2(\bbC), Sp(n))$ dual pair decomposition, we also classify all $Sp(n)$-equivariant operators on $\Lambda_{2n}$ that satisfy the fermionic Helmholtz relations and have fourth power equal to the identity. Here, the terminology ``Helmholtz relations'' and related characterizations of generalized Fourier transforms come from \cite{DOV16}.

In Section \ref{sec:hybrid}, we establish our main rigidity result for the hybrid Fourier transform in Theorem \ref{thm:generalFT}. For $m,n\ge1$, we show in Lemma \ref{lem:quadratic-oscillator} that the continuous ordinary commutant of the full even quadratic action $\frg_0 \cong \fso(4n)\oplus\fsp(2m)$ on $\caS(\bbR^{m|2n})$ is equal to
\[
 \bbC[\Pb,\Pf]
 = \spanC\{\Id,\Pb,\Pf,\Pb\Pf\} \mdot
\]
When we adjoin any one nonzero mixed quadratic relation from $\frg_1$, the space of intertwining operators reduces to the two-dimensional span
\[
 \spanC\{\Fhyb^+,\Fhyb^-\} \mdot
\]
Indeed, this reduction holds because the mixed quadratic space $\frg_1$ is an irreducible module for the even subalgebra $\frg_0$, so any nonzero mixed quadratic operator is cyclic under the adjoint action of $\frg_0$. In Theorem \ref{thm:generalFT}, we give both this single-relation characterization and the characterization via the explicit coordinate family of quadratic relations; imposing one first-order covariance relation then removes the remaining total-parity freedom. We compute the required commutants on $\caS(\bbR^{m|2n})$ by a Hermite--Grassmann basis argument, using continuity to pass from finite vectors to the full Schwartz--Grassmann space (cf. \cite{Nis90, C13} for the orthosymplectic oscillator representation and its parity decomposition).

In Section \ref{sec:gate}, we consider the unitary realization of the hybrid Fourier transform with respect to the occupation number inner product, in which the ordered Grassmann monomials form an orthonormal basis. In De Bie's original normalization, $\Gf^\sigma$ is not unitary in this inner product when $n \geq 1$. Therefore, we introduce the degree-scaling operator
\[
 \Dn \coloneqq 2^{-\bbE_f/2} \mcomma \qquad
 \Uf^\sigma \coloneqq \Dn^{-1}\Gf^\sigma\Dn \mcomma
\]
which gives the unitary normalization $\Uf^\sigma$ of $\Gf^\sigma$ in the occupation number inner product, and we define $\Uhyb^\sigma \coloneqq \Ub^\sigma\otimes\Uf^\sigma$ (cf. \cite{CD11, CDS11} for the corresponding weighted fermionic inner product). By transporting our quadratic rigidity theorem to the unitary setting, we show that any unitary intertwiner $T$ satisfying the quadratic relations has the form
\[
 T = e^{\compi\phi}e^{\compi\vartheta\Pb\Pf}\Uhyb^+ \mcomma
 \qquad \phi,\vartheta\in\bbR \mcomma
\]
and any single first-order canonical relation reduces $T$ to $e^{\compi\phi}\Uhyb^+$. Furthermore, for $m=1$ and $n\ge2$, if we impose full even quadratic covariance together with the fermion-parity preserving cubic relation for the density--displacement coupling $qQ_f$, we obtain $T = e^{\compi\phi}e^{\compi\theta\Pf}U_{1|2n}^+$ in Proposition \ref{corl:physical-number-identification}.

Up to a scalar phase, the normalized transform $\Uhyb^\sigma$ factors into a rotation in the bosonic oscillator representation and an $SU(2)$ rotation in the fermionic Lefschetz representation. On each symplectic pair of fermionic modes, the $4\times 4$ matrix of $\Uf^\sigma$ is a standard two-qubit matchgate (cf. \cite{JozsaMiyake2008}). In Sections \ref{subsec:density-quadrature} and 5.5, we apply the unitary conjugation formulas of $\Uhyb^\sigma$ to relate density--displacement and density--momentum Hamiltonians, conditional displacement gates, and two representations of the Hubbard--Holstein Hamiltonian on a finite chain. These applications connect our operator identities to hybrid qumode--qubit quantum simulation (cf. \cite{Andersen2015, Crane2026, Chen2025Genesis}). Note that the occupation-space unitary $\Uhyb^\sigma$ studied here is distinct from the Grassmann Fourier transforms used to construct fermionic and hybrid phase-space symbols in \cite{CahillGlauber1999, Sarkis2025}.

The paper is organized as follows. In Section \ref{sec:bosonic}, we recall the bosonic Fourier transform and the oscillator Howe decomposition. In Section \ref{sec:grassmann}, we study the fermionic Fourier transform on $\Lambda_{2n}$, its algebraic characterizations, and its decomposition under the dual pair $(\fsl_2(\bbC), Sp(n))$. In Section \ref{sec:hybrid}, we construct the hybrid Fourier transform $\Fhyb^\sigma$, examine its covariance properties, and prove the quadratic rigidity theorems. Finally, in Section \ref{sec:gate}, we give the unitary normalization $\Uhyb^\sigma$ in the occupation number basis and calculate its conjugation action on hybrid quantum gates and Hamiltonians. Throughout this paper, we set $\bbN = \{0,1,2,\ldots\}$ and $\compi = \sqrt{-1}$. The parameter $\sigma \in \{+1,-1\}$ denotes the sign in the Fourier kernel, so that $\Fhyb^-$ is the inverse of $\Fhyb^+$. Unless otherwise specified, all vector spaces, operator algebras, and representations are over $\bbC$.

\section{Bosonic Fourier transforms and Howe decomposition}\label{sec:bosonic}
In this section, we recall the definition of the bosonic Fourier transform, its coordinate and radial covariance properties, and the classical oscillator Howe decomposition.

\subsection{Bosonic Fourier transform and coordinate covariance}
\begin{defn}\cite{HZ19}\cite[Section 3]{DB08}
For $\sigma\in\{+1,-1\}$, the \textbf{bosonic Fourier transform} (for continuous variables) of $f \in \caS (\bbR^m)$ is defined by
\begin{align}
\begin{split}
\Ub^{\sigma} (f)(\bmy) & \coloneqq (2 \pi)^{- \frac{m}{2}} \int_{\bbR^m} \, \exp \left(\sigma \, \compi \, \bmx \cdot \bmy \right) f(\bmx)  \, d \bmx \\
& = (2 \pi)^{- \frac{m}{2}} \int_{\bbR^m} \, \exp \left(- \sigma \, \compi \, \langle \bmx, \bmy \rangle \right) f(\bmx) \, d \bmx \mdot
\end{split}
\end{align}
Here we use the negative-definite bilinear form $\langle\bmx,\bmy\rangle \coloneqq -\bmx\cdot\bmy$, so the two integral expressions above coincide. We take $\Ub^{+}$ as the bosonic Fourier transform and $\Ub^{-}$ as its inverse.
\end{defn}

\begin{prop}\label{prop:bosonicFourier1}\emph{\cite{HZ19}}
For each $\sigma\in\{+1,-1\}$, the operator $\Ub^{\sigma}$ has the following properties:
\begin{enumerate}[(1)]
\item The bosonic Fourier transform $\Ub^{\sigma}$ is a homeomorphism of the Schwartz space $\caS (\bbR^m)$ onto itself and extends to a unitary operator on $L^2(\bbR^m)$. Furthermore, $\Ub^{\sigma}$ commutes with the natural action of the orthogonal group $O(m)$ on $\caS (\bbR^m)$, namely, $\Ub^{\sigma}$ is $O(m)$-equivariant:
\[\Ub^{\sigma} (gf) = g (\Ub^{\sigma} \, f) \mcomma \quad \forall f \in \caS (\bbR^m) \mcomma \; \forall g \in O(m) \mdot\]
\item For $1 \leq r \leq m$, we have the following intertwining relations:
\begin{align*}
\Ub^{\sigma} \circ x_r &= - \sigma \, \compi \, \partial_{y_r} \circ \Ub^{\sigma} \mcomma \\
\Ub^{\sigma} \circ \partial_{x_r} & = - \sigma \, \compi \, y_r \circ \Ub^{\sigma} \mdot
\end{align*}
\item Let $\tilde{f}$ denote the reflection of $f$, i.e., $\tilde{f}(\bmx) = f(-\bmx)$ for all $\bmx \in \bbR^m$. Then we have
\[(\Ub^{\sigma})^2 f = \tilde{f} \mcomma \quad \forall f \in \caS (\bbR^m) \mdot\]
\end{enumerate}
\end{prop}

\begin{prop}\label{prop:bosonicFourier2}\emph{(cf.~\cite{HZ19}\cite[Propositions 1.2.1 and 1.2.2]{KM11})}
Suppose that $\caT$ is a continuous complex-linear operator on the Schwartz space $\caS (\bbR^m)$.
\begin{enumerate}[(1)]
\item For a fixed $\sigma\in\{+1,-1\}$, the condition
\[ \caT \circ x_r = - \sigma\,\compi\, \partial_{y_r} \circ \caT \quad \text{and} \quad \caT \circ \partial_{x_r} = - \sigma\,\compi\, y_r \circ \caT \quad (1 \leq r \leq m) \]
implies that $\caT$ is a scalar multiple of $\Ub^{\sigma}$.
\item The condition
\[ \caT \circ x_r x_s = - \partial_{y_r} \partial_{y_s} \circ \caT \quad \text{and} \quad \caT \circ \partial_{x_r} \partial_{x_s} = - y_r y_s \circ \caT \quad (1 \leq r, s \leq m) \]
implies that $\caT = a \; \Ub^{+} + b \; \Ub^{-}$ for some $a, b \in \bbC$.
\end{enumerate}
\end{prop}

Note that the corresponding $L^2(\bbR^m)$ statements are given in \cite[Propositions 1.2.1 and 1.2.2]{KM11}. By applying the same Hermite-basis argument together with continuity on $\caS(\bbR^m)$, we obtain the Schwartz-space statements in Proposition \ref{prop:bosonicFourier2}.

\subsection{Howe decomposition and radial covariance}

Let us define the radial differential operators on $\caS(\bbR^m)$ by
\[
 X_b \coloneqq \frac12|\bmx|^2 \mcomma \qquad
 Y_b \coloneqq -\frac12\Delta_{\mathrm E} \mcomma \qquad
 H_b \coloneqq \bbE_b+\frac m2 \mcomma \qquad \bbE_b \coloneqq \sum_{r=1}^m x_r\partial_{x_r} \mcomma
 \qquad \Delta_{\mathrm E} \coloneqq \sum_{r=1}^m\partial_{x_r}^2 \mdot
\]
By direct calculation, we see that these operators satisfy the $\fsl_2$-commutation relations $[X_b,Y_b]=H_b$, $[H_b,X_b]=2X_b$, and $[H_b,Y_b]=-2Y_b$, and they commute with the action of $O(m)$. Here $\Delta_{\mathrm E}$ is the standard Euclidean Laplacian; in Section \ref{sec:hybrid}, we will use the superspace sign convention $\Delta_b = -\Delta_{\mathrm E}$.
Put
\[
 \caH_k^b \coloneqq \ker\Delta_{\mathrm E}\cap\caP_k(\bbR^m) \mcomma \qquad
 \mathcal K_m \coloneqq \{k\geq0 :\, \caH_k^b\ne0\} \mcomma
\]
where $\caP_k(\bbR^m)$ denotes the space of homogeneous polynomials of degree $k$ on $\bbR^m$. Then we have $\mathcal K_1 = \{0,1\}$ and $\mathcal K_m = \bbN$ for $m \geq 2$.
For each $k \in \mathcal K_m$, let $\{H_{k,\ell}\}_\ell$ be a basis of the harmonic space $\caH_k^b$. Then the functions
\[
 \Phi_{j,k,\ell}(\bmx)
 \coloneqq L_j^{(k+m/2-1)}(|\bmx|^2)
 H_{k,\ell}(\bmx)e^{-|\bmx|^2/2} \mcomma \qquad j\geq0 \mcomma
\]
form a basis of the Hermite-finite subspace $\caS_{\rm fin}(\bbR^m)$, where $L_j^{(\alpha)}$ denotes the generalized Laguerre polynomial.
Under the joint action of $\fsl_2 \times O(m)$, we have the classical oscillator Howe decomposition (cf. \cite{How95, HZ19}):
\begin{equation}\label{eq:bosonic-howe-decomposition}
 \caS_{\rm fin}(\bbR^m)
 \cong \bigoplus_{k\in\mathcal K_m}
 D_{k+m/2}\otimes\caH_k^b \mcomma
\end{equation}
where $D_{k+m/2}$ is the irreducible lowest-weight $\fsl_2$-module with lowest compact weight $k+m/2$ relative to $X_b+Y_b$. On this radial string, the harmonic oscillator $X_b+Y_b$ acts with eigenvalues $2j+k+m/2$, and each harmonic space $\caH_k^b$ is an irreducible $O(m)$-module (including the case $m=2$).
Let $\widehat D_{k+m/2}$ denote the Hilbert space completion of $D_{k+m/2}$. Taking the Hilbert space direct sum, we get
\[
 L^2(\bbR^m) \cong
 \widehat{\bigoplus}_{k\in\mathcal K_m}
 \widehat D_{k+m/2}\widehat\otimes\caH_k^b \mdot
\]
Under our normalization of $\Ub^\sigma$, the basis functions $\Phi_{j,k,\ell}$ are eigenfunctions of $\Ub^\sigma$:
\begin{equation}\label{eq:bosonic-fourier-eigenvalues}
 \Ub^\sigma\Phi_{j,k,\ell}
 = (\sigma\compi)^{2j+k}\Phi_{j,k,\ell} \mdot
\end{equation}
Let us denote by $B_k^\sigma$ the radial action of $\Ub^\sigma$ on $D_{k+m/2}$ (as well as its unitary extension to $\widehat D_{k+m/2}$). Note that the algebraic direct sum in (\ref{eq:bosonic-howe-decomposition}) holds on the subspace of Hermite-finite vectors $\caS_{\rm fin}(\bbR^m)$, while general Schwartz functions in $\caS(\bbR^m)$ are infinite series in these components.

By using the Howe decomposition (\ref{eq:bosonic-howe-decomposition}), we obtain the following block diagonalization for operators satisfying radial covariance (cf. \cite{HZ19}).
\begin{prop}\label{prop:bosonic-block-principle}
Let $T$ be a bounded linear operator on $L^2(\bbR^m)$ that preserves $\caS_{\rm fin}(\bbR^m)$. Suppose that on $\caS_{\rm fin}(\bbR^m)$ we have
\[
 Tg = gT \quad (g\in O(m)) \mcomma \qquad
 TX_b = Y_bT \mcomma \qquad TY_b = X_bT \mdot
\]
Then, with respect to the Hilbert space decomposition above, $T$ has the block form
\begin{equation}\label{eq:bosonic-fourier-blocks}
 T \cong \widehat{\bigoplus}_{k\in\mathcal K_m}
 c_k\bigl(B_k^+\widehat\otimes\Id_{\caH_k^b}\bigr) \mcomma
 \qquad \sup_{k\in\mathcal K_m}|c_k| < \infty \mcomma
\end{equation}
where $B_k^+$ is the radial factor of $\Ub^+$ on $\widehat D_{k+m/2}$.
Conversely, any bounded sequence of scalars $\{c_k\}_{k\in\mathcal K_m}$ defines such an operator $T$.
Furthermore, $T$ satisfies $T^4 = \Id$ if and only if $c_k \in \{1,-1,\compi,-\compi\}$ for all $k \in \mathcal K_m$, and in this case $T$ is unitary.
\end{prop}

Note that any operator $T$ of the form (\ref{eq:bosonic-fourier-blocks}) also acts continuously on the Schwartz space $\caS(\bbR^m)$. Indeed, $T$ commutes with the harmonic oscillator $X_b+Y_b$, whose powers define the Fr\'echet topology of $\caS(\bbR^m)$ via their $L^2$-graph norms. Thus, the radial covariance conditions in Proposition \ref{prop:bosonic-block-principle} leave an independent scalar $c_k$ on each irreducible summand, whereas the stronger coordinatewise conditions in Proposition \ref{prop:bosonicFourier2} reduce the operator to a linear combination of $\Ub^+$ and $\Ub^-$.


\section{Fermionic Fourier transform on the exterior algebra}\label{sec:grassmann}
\subsection{Grassmann variables and Berezin integration}
Let $\Lambda_{2n}$ be the complex Grassmann algebra generated by $\grx_1,\ldots,\grx_{2n}$ subject to the anticommutation relations $\grx_a\grx_b = -\grx_b\grx_a$. For each ordered subset $I = \{i_1 < \cdots < i_k\} \subseteq \{1,\ldots,2n\}$, we write $\grx_I \coloneqq \grx_{i_1}\cdots\grx_{i_k}$, with $\grx_\emptyset \coloneqq 1$.
We define the left Grassmann derivative $\partial_{\grx_a}$ on $\Lambda_{2n}$ by
\[
 \partial_{\grx_a}(1) = 0 \mcomma \qquad
 \partial_{\grx_a}(\grx_b f) = \delta_{ab}f - \grx_b\partial_{\grx_a}f \mdot
\]
Then the left multiplication operators $\grx_a$ and contraction operators $\partial_{\grx_b}$ satisfy the canonical anticommutation relations $\{\grx_a,\partial_{\grx_b}\} = \delta_{ab}$, while all other anticommutators vanish. When we introduce a second set of Grassmann generators $\gry_1,\ldots,\gry_{2n}$, we require that $\grx_a$ and $\gry_b$ also anticommute with each other. We define the symplectic fermionic pairing by
\begin{equation}\label{eq:fermionic-pairing}
 \langle\bmgrx,\bmgry\rangle
 \coloneqq \frac12\sum_{j=1}^n
 (\grx_{2j-1}\gry_{2j}-\grx_{2j}\gry_{2j-1}) \mdot
\end{equation}
For any element $f = \sum_I f_I\grx_I \in \Lambda_{2n}$, its normalized Berezin integral is defined by (cf. \cite{Ber66, Ber87, DB08})
\begin{equation}\label{eq:fermionic-berezin}
 \int_{B,\bmgrx}f
 \coloneqq \pi^{-n}\partial_{\grx_{2n}}\cdots\partial_{\grx_1}f
 = \pi^{-n}f_{\{1,\ldots,2n\}} \mdot
\end{equation}
Here, the normalization factor $\pi^{-n}$ follows De Bie's convention in \cite{DB08}.

\subsection{Fermionic Fourier transform}
\begin{defn}\cite[Section 3]{DB08}
For $\sigma\in\{+1,-1\}$, the \textbf{fermionic Fourier transform} of $f \in \Lambda_{2n}$ is defined via the normalized Berezin integral with respect to the pairing (\ref{eq:fermionic-pairing}) by
\begin{align}\label{defn:fermionicFT}
\begin{split}
\Gf^{\sigma} (f)(\bmgry) & \coloneqq (2 \pi)^{n} \int_{B, \bmgrx} \exp \left(- \sigma \; \compi\; \langle \bmgrx, \bmgry \rangle \right) f(\bmgrx) \\
& = (2 \pi)^{n} \int_{B, \bmgrx} \exp \left(-\sigma \frac{\compi}{2} \sum_{j=1}^{n} (\grx_{2j-1} \grave{y}_{2j} - \grx_{2j} \grave{y}_{2j-1} )   \right) f(\bmgrx) \mdot
\end{split}
\end{align}
Since $\Lambda_{2n}$ is finite-dimensional, $\Gf^{\sigma}$ is a linear endomorphism of $\Lambda_{2n}$ (after identifying the $\bmgrx$-variables and $\bmgry$-variables). We take $\Gf^{+}$ as the fermionic Fourier transform and $\Gf^{-}$ as its inverse.
\end{defn}

\begin{prop}\emph{\cite[Section 3, Lemma 2]{DB08}}\label{prop:intertwining}
For any $f \in \Lambda_{2n}$ and $\sigma\in\{+1,-1\}$, we have the following intertwining relations:
\begin{align*}
\Gf^{\sigma}  (\partial_{\grx_{2j}} f) & = - \sigma \frac{\compi}{2} \gry_{2j -1} \Gf^{\sigma} (f) \mcomma\\
\Gf^{\sigma}  (\partial_{\grx_{2j-1}} f) & = + \sigma \frac{\compi}{2} \gry_{2j } \Gf^{\sigma} (f) \mcomma\\
\Gf^{\sigma}  (\grx_{2j} f) & = 2\sigma \compi \partial_{\gry_{2j -1}} \Gf^{\sigma} (f) \mcomma\\
\Gf^{\sigma}  (\grx_{2j-1} f) & = - 2\sigma \compi \partial_{\gry_{2j}} \Gf^{\sigma} (f) \mdot
\end{align*}
\end{prop}

Let us define the fermionic parity operator $P_f$ on $\Lambda_{2n}$ by $(P_f f)(\bmgrx) \coloneqq f(-\bmgrx)$. By \cite[Section 3, Theorem 3]{DB08}, we have the inversion and periodicity identities
\[
(\Gf^\sigma)^{-1} = \Gf^{-\sigma} \mcomma \qquad
(\Gf^\sigma)^2 = P_f \mcomma \qquad
P_f\Gf^\sigma = \Gf^{-\sigma} \mcomma \qquad
(\Gf^\sigma)^4 = \Id \mdot
\]

\subsection{Characterization of fermionic Fourier transform}

Now we use the intertwining relations in Proposition \ref{prop:intertwining} to characterize the fermionic Fourier transform on $\Lambda_{2n}$. When we write $\bmgrx$ for the input variables and $\bmgry$ for the output variables, we view $\caT$ as a linear map from the Grassmann algebra generated by $\bmgrx$ to the copy generated by $\bmgry$; under the canonical identification $\grx_a \leftrightarrow \gry_a$, we view $\caT$ as an endomorphism of $\Lambda_{2n}$.

\begin{thm}\label{thm:fermionicFT}
Suppose that $\caT$ is a linear operator on the Grassmann algebra $\Lambda_{2n}$.
\begin{enumerate}[(1)]
\item The condition
\[\begin{matrix*}[l] \caT  \circ \partial_{\grx_{2j}}   = - \frac{\compi}{2} \gry_{2j -1} \circ  \caT  \mcomma & \caT   \circ \partial_{\grx_{2j-1}}   =  \frac{\compi}{2} \gry_{2j } \circ  \caT \mcomma\\ \caT   \circ \grx_{2j}  =  2\compi \partial_{\gry_{2j -1}}  \circ \caT \mcomma & \caT   \circ \grx_{2j-1}   = - 2\compi \partial_{\gry_{2j}}  \circ \caT \mcomma \end{matrix*}  \quad (1 \leq j \leq n) \]
implies that $\caT$ is a scalar multiple of $\Gf^{+}$.
\item Assume that $n\geq2$. The condition
\[  \begin{matrix*}[l] \caT  \circ \partial_{\grx_{2j}} \partial_{\grx_{2k}}  = - \frac{1}{4} \gry_{2j -1} \gry_{2k -1} \circ  \caT  \mcomma &   \caT  \circ \partial_{\grx_{2j}} \partial_{\grx_{2k-1}}  =  \frac{1}{4} \gry_{2j -1} \gry_{2k } \circ  \caT \mcomma\\ \caT  \circ \partial_{\grx_{2j-1}} \partial_{\grx_{2k-1}}  =  -\frac{1}{4} \gry_{2j } \gry_{2k } \circ  \caT  \mcomma & \caT  \circ \partial_{\grx_{2j-1}} \partial_{\grx_{2k}}  =  \frac{1}{4} \gry_{2j } \gry_{2k-1 } \circ  \caT \mcomma\\ \caT \circ \grx_{2j} \grx_{2k} = -4 \partial_{\gry_{2j-1}} \partial_{\gry_{2k-1}} \circ \caT \mcomma & \caT \circ \grx_{2j} \grx_{2k-1} = 4 \partial_{\gry_{2j-1}} \partial_{\gry_{2k}} \circ \caT \mcomma\\ \caT \circ \grx_{2j-1} \grx_{2k-1} = -4 \partial_{\gry_{2j}} \partial_{\gry_{2k}} \circ \caT \mcomma & \caT \circ \grx_{2j-1} \grx_{2k} = 4 \partial_{\gry_{2j}} \partial_{\gry_{2k-1}} \circ \caT \mcomma \end{matrix*}  \quad (1 \leq j,\, k\, \leq n)
\]
implies that $\caT = a \; \Gf^{+} + b \; \Gf^{-}$ for some $a, \, b \in \bbC$.
\end{enumerate}
For $n=0$, both statements hold trivially since $\Lambda_0 \cong \bbC$.
\end{thm}

\begin{proof}
\textbf{(1)}
Let $V$ be a $4n$-dimensional complex vector space equipped with a nondegenerate symmetric bilinear form $\kappa$. Let $\{z_1,\, z_2, \, \ldots , \, z_{4n}\}$ be an orthonormal basis of $V$ with respect to $\kappa$. In the Clifford algebra $C\ell(V,\kappa)$, these basis elements satisfy the relations
$$z_j z_k + z_k z_j = 2 \kappa(z_j, z_k) = 2 \delta_{jk} \mdot$$
We choose the Witt basis (cf. \cite{HP06}):
\begin{equation}
f_j \coloneqq \frac{z_{2j -1} + \compi z_{2j}}{2} \mcomma \qquad f_{j}^{*} \coloneqq \frac{z_{2j -1} - \compi z_{2j}}{2} \mcomma \qquad j=1,\, \ldots,\, 2n \mdot
\end{equation}
For $1\leq j, \, k \leq 2n$, we have $\kappa (f_j , f_k^*) = \frac{1}{2} \delta_{jk}$, and we get the anticommutation relations
\begin{align*}
f_j f_k + f_k f_j &= 0 \mcomma \\
f_j^* f_k^* + f_k^* f_j^* &= 0 \mcomma\\
f_j f_k^* + f_k^* f_j &= \delta_{jk} \mdot
\end{align*}
Let $U$ be the subspace spanned by $\{f_j :\; j=1,\,\ldots,\, 2n\}$, and let $U^*$ be the subspace spanned by $\{f_j^* :\; j=1,\,\ldots,\, 2n\}$. Then $U$ and $U^*$ form a pair of complementary $2n$-dimensional isotropic subspaces of $V$ paired nondegenerately by $\kappa$, so we have
\begin{equation}
V = U \oplus U^* = \vectorspan\{f_1, f_2, \ldots ,f_{2n}\} \oplus \vectorspan\{f_1^* , f_2^*, \ldots , f_{2n}^*\} \mdot
\end{equation}

On the Grassmann algebra $\Lambda_{2n}$, the multiplication and contraction operators
\[
\{ \grave{x}_j, \partial_{\grave{x}_j} \mid 1 \leq j \leq 2n \}
\]
satisfy the canonical anticommutation relations and generate an associative algebra isomorphic to $C\ell(V,\kappa)$ via the correspondence
$$ \grave{x}_j \longmapsto f_j \mcomma \qquad \partial_{\grave{x}_j} \longmapsto f_j^* \mdot$$
We list the relations between these generators in Table \ref{table:LieSuperCliffordAlg}.
\begin{table}[htb]
  \centering
\begin{tabular}{ c : c }
\hline
Grassmann operators & Clifford algebra  \\
\hline
$\grx_j \grx_k + \grx_k \grx_j = 0 $ & $f_j f_k + f_k f_j = 0$ \\
$\partial_{\grx_j} \partial_{\grx_k} + \partial_{\grx_k} \partial_{\grx_j} = 0 $ &  $f_j^* f_k^* + f_k^* f_j^* = 0$\\
$\grx_j \partial_{\grx_k} + \partial_{\grx_k} \grx_j = \delta_{jk} $ & $f_j f_k^* + f_k^* f_j = \delta_{jk}$\\
\hline
\end{tabular}
\caption{Correspondence between generators} \label{table:LieSuperCliffordAlg}
\end{table}

We identify the Grassmann algebra $\caP = \Lambda_{2n}$ with the exterior algebra $\bigwedge U$, which is the unique irreducible spin module for the Clifford algebra $C\ell(V, \kappa)$ (cf. \cite{HP06}). Thus, $\caP$ is a simple module over the associative algebra $C\ell(V, \kappa)$.

The conditions in (1) describe how $\caT$ intertwines the generators $\{\grx_j, \gry_j, \partial_{\grx_j}, \partial_{\gry_j} \}$. By Proposition \ref{prop:intertwining}, we know that $\Gf^{+}$ satisfies the exact same first-order intertwining relations as $\caT$. Let us put
\[
\widetilde{\caT} \coloneqq \caT \circ (\Gf^{+})^{-1} = \caT \circ \Gf^{-} \mdot
\]
Then $\widetilde{\caT}$ commutes with every multiplication and contraction operator $\gry_j$ and $\partial_{\gry_j}$. Indeed, from $\Gf^{+}\circ \partial_{\grx_{2j-1}} = \frac{\compi}{2}\,\gry_{2j}\circ \Gf^{+}$, we have $\gry_{2j} = -2\compi\,\Gf^{+}\circ \partial_{\grx_{2j-1}}\circ \Gf^{-}$, and hence we get
\[
\widetilde{\caT} \circ \gry_{2j}
= \caT \circ \Gf^{-} \circ \gry_{2j}
= -2\compi\, \caT \circ \partial_{\grx_{2j-1}} \circ \Gf^{-}
= -2\compi \cdot \tfrac{\compi}{2}\, \gry_{2j} \circ \caT \circ \Gf^{-}
= \gry_{2j} \circ \widetilde{\caT} \mdot
\]
We can check the remaining generators $\gry_{2j-1}$, $\partial_{\gry_{2j}}$, and $\partial_{\gry_{2j-1}}$ in the same way.

Since $\{\gry_j, \partial_{\gry_j} \mid 1\le j\le 2n\}$ generate the associative Clifford algebra $C\ell(V,\kappa)$, and $\caP = \Lambda_{2n}$ is an irreducible module over $C\ell(V,\kappa)$, Burnside's theorem implies that the image of $C\ell(V,\kappa)$ is the full matrix algebra $\End(\caP)$. Since $\widetilde{\caT}$ commutes with all generators of $\End(\caP)$, by Schur's lemma, we have $\widetilde{\caT} = c\Id$ for some $c\in\bbC$. Therefore, we get $\caT = c\,\Gf^{+}$.

\textbf{(2)}
By Lemma \ref{lem:pure-quadratics-generate} below, the quadratic operators $\{ \grave{x}_j\grave{x}_k, \partial_{\grave{x}_j}\partial_{\grave{x}_k}, \grave{x}_j\partial_{\grave{x}_k} - \frac{1}{2} \delta_{jk} \}$ span the orthogonal Lie algebra $\fso(V, \bbC)=\Lambda^2U\oplus\fgl(U)\oplus\Lambda^2U^*$ in its spin representation.
Let us define $h_j \coloneqq f_j^* f_j - \frac{1}{2}$ for $j=1, 2, \ldots , 2n$, and let $\fh$ denote the subspace spanned by $\{h_j :\, 1 \leq j \leq 2n\}$. Then $\fh$ is a Cartan subalgebra of $\fso(V, \bbC)$ of rank $2n$. We list the roots and root vectors of $\fso(V, \bbC)$ with respect to $\fh$ in Table \ref{table:rootvectors} (cf. \cite{HP06}), where the roots are given by
\[
\alpha_{jk}(h_l) = \delta_{jl}-\delta_{kl}\ \ (j\ne k) \mcomma
\qquad
\beta_{jk}(h_l) = \delta_{jl}+\delta_{kl}\ \ (j<k) \mdot
\]
\begin{table}[htb]
  \centering
\begin{tabular}{ c : c }
\hline
root & corresponding root vector  \\
\hline
$\alpha_{jk}$ for $j \ne k$ & $f_j^* f_k$ \\
$-\beta_{jk}$ for $j < k$ & $f_j f_k$ \\
$\beta_{jk}$ for $j < k$ & $f_j^* f_k^*$ \\
\hline
\end{tabular}
\caption{Roots and root vectors of $\fso(V, \bbC)$} \label{table:rootvectors}
\end{table}

The spin representation of $\fso(V, \bbC)$ on $\caP$ is given by the correspondence of quadratic generators:
\begin{align*}
\grx_j \grx_k &\longmapsto f_j f_k \mcomma \\
\partial_{\grx_j} \partial_{\grx_k} & \longmapsto f_j^* f_k^* \mcomma \\
\grx_j \partial_{\grx_k} - \tfrac12\delta_{j k}\Id & \longmapsto f_j f_k^* - \tfrac12\delta_{jk} \mdot
\end{align*}
In fact, this Lie algebra is the image of $\bigwedge^2 V$ under the Chevalley map (cf. \cite{HP06}).
To describe the action of $\fso(V, \bbC)$ on $\caP$, let $\bbE$ be the fermionic Euler operator defined by
$$ \bbE \coloneqq \sum_{j=1}^{2n} \grave{x}_j \partial_{\grave{x}_j} \mdot $$
Using $\bbE$, we get a natural degree grading on the Grassmann algebra $\caP = \Lambda_{2n}$. Let $\caP_k$ denote the eigenspace of $\bbE$ with eigenvalue $k$, which is the subspace of homogeneous Grassmann polynomials of degree $k$:
$$ \caP_k \coloneqq \{ \omega \in \caP \mid \bbE\omega = k\omega \} \mdot $$
Then we have the direct sum decomposition
$$ \caP = \bigoplus_{k=0}^{2n} \caP_k \mdot $$

Since all quadratic operators in $\fso(V, \bbC)$ have even degree, their action preserves the parity of the degree $k$. Consequently, the spin module $\caP$ decomposes into two inequivalent irreducible half-spin representations of $\fso(V, \bbC)$:
$$ \caP = \caP_{\text{even}} \oplus \caP_{\text{odd}} \mcomma $$
where the even and odd subspaces are defined by
$$ \caP_{\text{even}} \coloneqq \bigoplus_{k \geq 0} \caP_{2k} \mcomma \qquad \caP_{\text{odd}} \coloneqq \bigoplus_{k \geq 0} \caP_{2k+1} \mdot $$

The conditions in (2) mean that $\caT$ intertwines the pure quadratic generators in $\Lambda^2U$ and $\Lambda^2U^*$. Since $\Gf^{+}$ satisfies the same quadratic intertwining relations, the operator $\widetilde{\caT} = \caT \circ (\Gf^{+})^{-1}$ commutes with all elements in $\Lambda^2U$ and $\Lambda^2U^*$. When $n\ge2$, by Lemma \ref{lem:pure-quadratics-generate}, $\Lambda^2U$ and $\Lambda^2U^*$ generate the full Lie algebra $\fso(V, \bbC)$. Hence $\widetilde{\caT}$ commutes with the action of $\fso(V, \bbC)$ on $\caP$.

Since $\caP_{\text{even}}$ and $\caP_{\text{odd}}$ are inequivalent irreducible $\fso(V, \bbC)$-modules, by Schur's lemma, there exist scalars $c_1, c_2 \in \bbC$ such that
$$ \widetilde{\caT}\big|_{\caP_{\text{even}}} = c_1 \Id \quad \text{and} \quad \widetilde{\caT}\big|_{\caP_{\text{odd}}} = c_2 \Id \mdot $$
Note that $(\Gf^{+})^{-1} \circ \Gf^{-} = P_f$ acts as $\Id$ on $\caP_{\text{even}}$ and as $-\Id$ on $\caP_{\text{odd}}$. Putting $a = \frac{c_1 + c_2}{2}$ and $b = \frac{c_1 - c_2}{2}$, we get $\widetilde{\caT} = a\Id + b P_f$, and hence
$$ \caT = a \Gf^{+} + b \Gf^{-} \mdot $$
This finishes the proof.
\end{proof}

\begin{lemma}\label{lem:pure-quadratics-generate}
Let $U=\vectorspan\{f_1,\ldots,f_{2n}\}$, $U^*=\vectorspan\{f_1^*,\ldots,f_{2n}^*\}$, and $V=U\oplus U^*$ be as above, and identify $\{\grx_j\}\leftrightarrow\{f_j\}$ and $\{\partial_{\grx_j}\}\leftrightarrow\{f_j^*\}$.
\begin{enumerate}[(1)]
\item If $n\ge2$, the Lie algebra generated by the pure creation quadratics $\{\grx_j\grx_k\}=\Lambda^2U$ and the pure annihilation quadratics $\{\partial_{\grx_j}\partial_{\grx_k}\}=\Lambda^2U^*$ is the full orthogonal Lie algebra $\fso(V,\bbC)=\Lambda^2U\oplus\fgl(U)\oplus\Lambda^2U^*$ in its spin representation. When $n=1$, these pure quadratics generate only a proper subalgebra isomorphic to $\fsl_2(\bbC)$.
\item The centered operators $\{\grx_j\partial_{\grx_k}-\tfrac12\delta_{jk}\Id\}$ span $\fgl(U)$, and the spin embedding of $\fso(V,\bbC)$ into the Clifford algebra is given by
\[
\grx_j\grx_k \longmapsto f_jf_k \mcomma
\qquad
\partial_{\grx_j}\partial_{\grx_k} \longmapsto f_j^*f_k^* \mcomma
\qquad
\grx_j\partial_{\grx_k}-\tfrac12\delta_{jk}\Id \longmapsto f_jf_k^*-\tfrac12\delta_{jk} \mdot
\]
\end{enumerate}
\end{lemma}

\begin{proof}
Put $d = 2n$. When $n \geq 2$, we have $d \geq 4$, so for any $a \neq b$ we can pick an auxiliary index $c \notin \{a,b\}$. Using the commutator identity $[AB,CD]=A\{B,C\}D-AC\{B,D\}+\{A,C\}DB-C\{A,D\}B$ and the canonical anticommutation relations, we calculate directly that
\[
[\grx_a\grx_c,\ \partial_{\grx_c}\partial_{\grx_b}]
= \grx_a\partial_{\grx_b} \mdot
\]
For $a\ne b$, this gives the off-diagonal generator $\grx_a\partial_{\grx_b} = \grx_a\partial_{\grx_b}-\tfrac12\delta_{ab}\Id$ of $\fgl(U)$. For $a=b$, the same commutator gives
\[
[\grx_a\grx_c,\ \partial_{\grx_c}\partial_{\grx_a}]
= \grx_a\partial_{\grx_a}+\grx_c\partial_{\grx_c}-1 \mdot
\]
Taking the difference of two such brackets for distinct auxiliary indices $c\ne c'$, we get $\grx_c\partial_{\grx_c}-\grx_{c'}\partial_{\grx_{c'}}$. Combining this with $\grx_a\partial_{\grx_a}+\grx_c\partial_{\grx_c}-1$, we obtain each centered diagonal generator $\grx_a\partial_{\grx_a}-\tfrac12\Id$. Thus we generate the whole subalgebra $\fgl(U)$, and together with $\Lambda^2U$ and $\Lambda^2U^*$ we get the full Lie algebra $\fso(V,\bbC)$.

On the other hand, when $n=1$ (so $d=2$), there is no third index $c$: both $\Lambda^2U = \bbC \grx_1\grx_2$ and $\Lambda^2U^* = \bbC \partial_{\grx_1}\partial_{\grx_2}$ are one-dimensional, and their commutator $[\grx_1\grx_2, \partial_{\grx_2}\partial_{\grx_1}] = \grx_1\partial_{\grx_1} + \grx_2\partial_{\grx_2} - 1$ spans only a one-dimensional Cartan subspace. Hence they generate only a three-dimensional subalgebra $\fsl_2(\bbC) \subset \fso_4(\bbC) \cong \fsl_2(\bbC)\oplus\fsl_2(\bbC)$. This finishes the proof.
\end{proof}

\begin{remark}
When $n=1$, the pure creation and annihilation quadratics $\grx_1\grx_2$ and $\partial_{\grx_1}\partial_{\grx_2}$ act irreducibly on $\caP_{\mathrm{even}}=\vectorspan\{1,\grx_1\grx_2\}$ and act by zero on $\caP_{\mathrm{odd}}=\vectorspan\{\grx_1,\grx_2\}$. Therefore, an operator $\caT \in \End(\Lambda_2)$ satisfies the pure quadratic relations in Theorem \ref{thm:fermionicFT}(2) if and only if
\[
 \caT = (c\Id_{\caP_{\mathrm{even}}}\oplus B)\Gf^+ \mcomma
 \qquad c\in\bbC \mcomma \quad B\in\End(\caP_{\mathrm{odd}}) \mcomma
\]
so the space of solutions has dimension $1 + 4 = 5$.
\end{remark}

\subsection{The $(\fsl_2,\mathrm{Sp}(n))$ dual pair decomposition}

In this subsection, we use the symplectic action on $U$ to classify $Sp(n)$-equivariant operators satisfying radial intertwining relations. We keep the notations $\caP = \Lambda_{2n} = \bigwedge U$, $\caP_r$, and $\bbE$ from the previous subsection.
We define the symplectic quadratic invariant $\bm{\grave{x}}^{2}$ and the fermionic Laplacian $\Delta_f$ by
\begin{align*}
\bm{\grave{x}}^{2} &\coloneqq \sum_{j=1}^n \grave{x}_{2j-1}\grave{x}_{2j} \mcomma \\
\Delta_f &\coloneqq 4 \sum_{j=1}^n \partial_{\grx_{2j-1}} \partial_{\grx_{2j}} \mdot
\end{align*}

We regard $U = \vectorspan\{\grx_1, \ldots, \grx_{2n}\}$ as a $2n$-dimensional complex symplectic vector space equipped with the alternating bilinear form whose matrix in this basis is the skew coefficient matrix in (\ref{eq:fermionic-pairing}). Here and henceforth, we denote by $Sp(2n,\bbC)=Sp(U)$ the complex symplectic group and by $Sp(n)$ its compact real form. The natural action of $Sp(2n,\bbC)$ on $U$ extends to an action by algebra automorphisms on $\caP=\bigwedge U$, and the contragredient action on $U^*$ corresponds to the action on the contraction operators $\partial_{\grx_a}$. Since both the bivector $\bm{\grave{x}}^2$ and the differential operator $\Delta_f$ are $Sp(n)$-invariant, we see that the three operators
$$X \coloneqq \frac{1}{2}\bm{\grave{x}}^2 \mcomma \quad Y \coloneqq -\frac{1}{2}\Delta_f \mcomma \quad \text{and} \quad H \coloneqq \bbE - n $$
commute with the action of $Sp(n)$ on $\caP$.

By direct calculation of their commutators, we get the standard $\fsl_2(\bbC)$-relations:
\[
[X,Y] = H \mcomma \qquad [H,X] = 2X \mcomma \qquad [H,Y] = -2Y \mdot
\]
Thus, the operators $X$, $Y$, and $H$ span a Lie algebra $\fs = \langle X, Y, H \rangle$.
Recall the standard basis of $\fsl_2(\bbC)$:
\begin{equation}\label{eq:sl2c_basis}
x=\begin{pmatrix} 0 & 1 \\ 0 & 0 \end{pmatrix} \mcomma \quad y=\begin{pmatrix} 0 & 0 \\ 1 & 0 \end{pmatrix} \mcomma \quad h=\begin{pmatrix} 1 & 0 \\ 0 & -1 \end{pmatrix}\mdot
\end{equation}
We define the Lie algebra homomorphism
\[ \omega :\, \fsl_2(\bbC) \longrightarrow \fs \]
by $\omega (x) = X$, $\omega (y)=Y$, and $\omega (h) = H$. When $n\ge1$, $\omega$ is a Lie algebra isomorphism from $\fsl_2(\bbC)$ onto $\fs$; when $n=0$, $\omega$ is the trivial representation on $\Lambda_0=\bbC$ (in which case our decomposition formulas below contain only the $k=0$ term).

Since the action of $Sp(n)$ on $\caP$ commutes with $\fs = \omega(\fsl_2(\bbC))$, $\caP$ is a joint $\fsl_2(\bbC) \times Sp(n)$-module. We recall the classical symplectic Lefschetz (or dual pair) decomposition of $\caP$ as follows.

\begin{lemma} \emph{\cite{DBES09} \cite{GW98}} \label{lem:dual_pair}
For each $k \in \{0, \dots, n\}$, let
\[
\caH_k^f \coloneqq \{\phi \in \caP_k :\, \Delta_f (\phi) = 0\}
\]
denote the space of fermionic spherical harmonics of degree $k$. Then $\caH_k^f$ is an irreducible $Sp(n)$-module. Furthermore, under the joint action of the dual pair $\left(\fsl_2(\bbC), Sp(n)\right)$, we have the multiplicity-free decomposition
\[
\caP \cong \bigoplus_{k=0}^n V_{n-k} \otimes \caH_k^f \mcomma
\]
where $V_{n-k}$ is the irreducible $\fsl_2(\bbC)$-module of dimension $n-k+1$ (with highest weight $n-k$), generated by applying the raising operator $X$ repeatedly to lowest-weight vectors in $\caH_k^f$.
\end{lemma}

\begin{remark}
Equivalently, for any nonzero harmonic polynomial $\phi \in \caH_k^f$, the $\fsl_2(\bbC)$-cyclic subspace
\[
\langle \phi \rangle_{\fsl_2} = \spanC\{\phi, X\phi, \dots, X^{n-k}\phi\}
\]
is an irreducible $\fsl_2(\bbC)$-module isomorphic to $V_{n-k}$.
\end{remark}

To write down this decomposition explicitly, we fix a basis $\{H_k^{(j)} :\, 1 \le j \le \dim \caH_k^f\}$ of $\caH_k^f$. For each pair $(k,j)$, we define
\[
W_k^{(j)} \coloneqq \spanC \bigl\{ (\bm{\grave{x}}^{\,2})^\ell H_k^{(j)} :\, 0 \le \ell \le n-k \bigr\} \mdot
\]
Then $W_k^{(j)}$ is an irreducible $\fsl_2(\bbC)$-module isomorphic to $V_{n-k}$ with lowest-weight vector $H_k^{(j)}$.

\begin{lemma}\label{lemma:sl2strings}
For each fixed degree $k \in \{0, \dots, n\}$ and index $j \in \{1, \dots, \dim\caH_k^f\}$, the weight spaces of the irreducible $\fsl_2(\bbC)$-module $W_k^{(j)}$ are the one-dimensional subspaces
\[
\bbC\,(\bm{\grave{x}}^{\,2})^\ell H_k^{(j)} \mcomma \qquad 0 \le \ell \le n-k \mdot
\]
Up to nonzero scalar factors, the operators $X$ and $Y$ map adjacent weight spaces bijectively onto one another:
\[
H_k^{(j)}
\;\;\stackarrows{Y}{X}\;\;
\bm{\grave{x}}^{\,2}H_k^{(j)}
\;\;\stackarrows{Y}{X}\;\;
\cdots
\;\;\stackarrows{Y}{X}\;\;
(\bm{\grave{x}}^{\,2})^{n-k}H_k^{(j)} \mdot
\]
Furthermore, we have $W_k^{(j)} \cong W_p^{(i)}$ as $\fsl_2(\bbC)$-modules if and only if $k = p$.
\end{lemma}

\begin{proof}
Since $H_k^{(j)} \in \caH_k^f = \ker(Y)\cap\caP_k$, we have $Y H_k^{(j)} = 0$ and $H H_k^{(j)} = (k-n)H_k^{(j)}$. Thus $H_k^{(j)}$ is a lowest-weight vector of weight $k-n$ for $\fsl_2(\bbC)$. By the standard representation theory of $\fsl_2(\bbC)$, applying $X$ successively produces a weight string of length $n-k+1$ and spans an irreducible module $W_k^{(j)} \cong V_{n-k}$. Finally, two finite-dimensional irreducible $\fsl_2(\bbC)$-modules $W_k^{(j)}$ and $W_p^{(i)}$ are isomorphic if and only if they have the same dimension, which means $n-k+1 = n-p+1$, i.e., $k=p$. This finishes the proof.
\end{proof}

Taking the direct sum of $W_k^{(j)}$ over all $k$ and $j$, we get the decomposition
\[
\caP = \bigoplus_{k=0}^n \bigoplus_{j=1}^{\dim \caH_k^f} W_k^{(j)} \cong \bigoplus_{k=0}^n V_{n-k} \otimes \caH_k^f \mdot
\]
In particular, comparing the dimensions of the homogeneous components gives the dimension formula for $\caH_k^f$:
\[
\dim \caH_k^f = \binom{2n}{k} - \binom{2n}{k-2}\mcomma
\]
where we set $\binom{2n}{r}=0$ for $r<0$.

\begin{corl}
Multiplication by powers of $\bm{\grave{x}}^2$ gives an $Sp(n)$-equivariant linear isomorphism
\begin{align*}
\bigoplus_{k=0}^n \bbC[\bm{\grave{x}}^{\,2}]_{\le n-k} \otimes \caH_k^f & \xrightarrow{\sim} \quad \caP \\
f(\bm{\grave{x}}^{\,2}) \otimes \phi \;\; &\longmapsto f(\bm{\grave{x}}^{\,2})\phi \mcomma
\end{align*}
where $\bbC[\bm{\grave{x}}^{\,2}]_{\le n-k}$ denotes the space of polynomials in $\bm{\grave{x}}^{\,2}$ of degree at most $n-k$. Equivalently, for each degree $0\leq r\leq2n$, the homogeneous component $\caP_r$ decomposes as the direct sum
\[
\caP_r = \bigoplus_{p=\max(0,r-n)}^{\lfloor \frac{r}{2} \rfloor} (\bm{\grave{x}}^{\,2})^p \caH_{r-2p}^f \mcomma
\]
where $\lfloor \frac{r}{2} \rfloor$ denotes the greatest integer less than or equal to $\frac{r}{2}$.
\end{corl}

\begin{proof}
By Lemma \ref{lemma:sl2strings}, each basis element $H_k^{(j)}$ generates an irreducible $\fsl_2(\bbC)$-module $W_k^{(j)} \cong V_{n-k}$. Taking the direct sum over $k$ and $j$, we immediately get the first isomorphism.

For the homogeneous subspace $\caP_r$ of degree $r$, any summand of total degree $r$ has the form $(\bm{\grave{x}}^{\,2})^p \phi$, where $\phi \in \caH_{k}^f$ and $k = r - 2p$. This summand is nonzero if and only if the harmonic degree satisfies $0 \leq k \leq n$ and the power $p$ satisfies $0 \leq p \leq n - k$. Substituting $k = r-2p$ into $k \geq 0$ and $p \leq n - k$, we get $2p \leq r$ and $p \geq r - n$, namely $\max(0, r-n) \le p \le \lfloor \frac{r}{2} \rfloor$. Combining the above calculations, we get our result.
\end{proof}

\subsection{Decomposition of fermionic Fourier transform}

Now we calculate the action of the fermionic Fourier transform $\Gf^\sigma$ on the dual pair decomposition in Lemma \ref{lem:dual_pair}. For each $k \in \{0, \dots, n\}$, let $\{H_{k,l}^f :\, 1 \le l \le \dim \caH_k^f\}$ be a basis of $\caH_k^f$. We define
\[
\widetilde{H}_{k,l}^f \coloneqq H_{k,l}^f \exp\left(\frac{\bm{\grave{x}}^{\,2}}{2}\right) \mdot
\]
Note that the exponential series above terminates as a finite sum because $\bm{\grave{x}}^2$ is nilpotent in $\Lambda_{2n}$. Next, let us put
\begin{align*}
\omega(K) &\coloneqq \frac{1}{2}(\Delta_f - \bm{\grave{x}}^{\,2}) \mcomma \\
\omega(E_+) &\coloneqq \frac{1}{4}(\Delta_f + \bm{\grave{x}}^{\,2} + 2\bbE - 2n) \mcomma \\
\omega(E_-) &\coloneqq \frac{1}{4}(-\Delta_f - \bm{\grave{x}}^{\,2} + 2\bbE - 2n) \mcomma
\end{align*}
which are the images under $\omega$ of the $\fsl_2(\bbC)$-basis elements
\begin{align*}
K & = -x - y = \begin{pmatrix} 0 & -1 \\ -1 & 0 \end{pmatrix} \mcomma \\
E_+ & = \frac{1}{2} (-y + x + h) = \frac{1}{2}\begin{pmatrix} 1 & 1 \\ -1 & -1 \end{pmatrix} \mcomma \\
E_- & = \frac{1}{2} (y - x + h) = \frac{1}{2}\begin{pmatrix} 1 & -1 \\ 1 & -1 \end{pmatrix} \mdot
\end{align*}

\begin{lemma}\label{lem:clifford_hermite_basis}
For each radial index $0 \le j \le n-k$, we define the Clifford--Hermite basis element by
\[
\psi_{j,k,l} \coloneqq (4\omega(E_+))^j \widetilde{H}_{k,l}^f \mdot
\]
Then we have the following properties:
\begin{enumerate}[(1)]
    \item For each fixed pair $(k,l)$, the subspace $\Psi_{k,l} \coloneqq \spanC\{\psi_{j,k,l} :\, 0 \le j \le n-k\}$ is an irreducible $\fsl_2(\bbC)$-module isomorphic to $V_{n-k}$.
    \item For each fixed $k$ and $j$, the set $\{\psi_{j,k,l} :\, 1 \le l \le \dim \caH_k^f\}$ forms a basis of an irreducible $Sp(n)$-module isomorphic to $\caH_k^f$.
    \item The basis elements $\psi_{j,k,l}$ satisfy the relations
    \begin{align*}
        \omega(K)\psi_{j,k,l} &= (2j + k - n)\psi_{j,k,l} \mcomma \\
        \omega(E_+)\psi_{j,k,l} &= \frac{1}{4}\psi_{j+1,k,l} \mcomma \\
        \omega(E_-)\psi_{j,k,l} &= -2j(2j - 2 - 2n + 2k)\psi_{j-1,k,l} \mcomma
    \end{align*}
    where we set $\psi_{-1,k,l} = \psi_{n-k+1,k,l} = 0$.
\end{enumerate}
\end{lemma}

\begin{proof}
Under the representation $\omega$, multiplication by $\exp(\bm{\grave{x}}^2/2)$ is equal to the operator $\exp(X)$. Using the commutation relations $[\omega(K),X]=H$, $[H,X]=2X$, $[\omega(E_-),X]=X-\tfrac12H$, and $[[\omega(E_-),X],X]=-X$, we calculate by the Baker--Campbell--Hausdorff formula that
\[
e^{-X}\omega(K)e^{X} = \omega(K)+[\omega(K),X]+\tfrac12\bigl[[\omega(K),X],X\bigr] = \omega(K)+H+X = H-Y \mcomma
\]
and
\[
e^{-X}\omega(E_-)e^{X} = \omega(E_-)+[\omega(E_-),X]+\tfrac12\bigl[[\omega(E_-),X],X\bigr] = \omega(E_-)+X-\tfrac12H-\tfrac12X = \tfrac12Y \mdot
\]
Since $H_{k,l}^f \in \caH_k^f$ satisfies $YH_{k,l}^f=0$ and $HH_{k,l}^f=(k-n)H_{k,l}^f$, we get
\[
\omega(E_-)\widetilde H_{k,l}^f = 0 \mcomma
\qquad
\omega(K)\widetilde H_{k,l}^f = (k-n)\widetilde H_{k,l}^f \mdot
\]
This shows that $\widetilde H_{k,l}^f$ is a lowest-weight vector of weight $k-n$ with respect to the $\fsl_2$-triple $(\omega(K),\omega(E_+),\omega(E_-))$, which satisfies
\[
[\omega(K),\omega(E_+)] = 2\omega(E_+) \mcomma
\qquad
[\omega(K),\omega(E_-)] = -2\omega(E_-) \mcomma
\qquad
[\omega(E_+),\omega(E_-)] = \omega(K) \mdot
\]
From the definition $\psi_{j,k,l}=(4\omega(E_+))^j\widetilde H_{k,l}^f$ and the standard $\fsl_2(\bbC)$ representation formulas, we immediately obtain the action of $\omega(K)$, $\omega(E_+)$, and $\omega(E_-)$ in (3), as well as (1). Finally, since both $\exp(\bm{\grave{x}}^2/2)$ and $\omega(E_+)$ commute with the action of $Sp(n)$, we see that for each fixed $j$ and $k$, the span of $\{\psi_{j,k,l}\}_l$ is an irreducible $Sp(n)$-module isomorphic to $\caH_k^f$. This finishes the proof.
\end{proof}

\begin{remark}
The subspace
\[
\tilde{\caH}_{k}^{f} \coloneqq \caH_{k}^{f} \exp\left(\frac{\bm{\grave{x}}^{\,2}}{2}\right)
\]
has basis $\{\psi_{0,k,l} :\, 1\le l\le\dim\caH_k^f\}$. It is the lowest-weight space with respect to $(\omega(K), \omega(E_+), \omega(E_-))$ and is isomorphic to $\caH_k^f$ as an $Sp(n)$-module.
\end{remark}

\begin{prop}\label{prop:fourier_decomp}
For each $\sigma\in\{+1,-1\}$, under the dual pair decomposition
\[
\caP \cong \bigoplus_{k=0}^n V_{n-k}\otimes \caH_k^f \mcomma
\]
the fermionic Fourier transform $\Gf^{\sigma}$ acts only on the radial factor $V_{n-k}$ and acts as the identity on the harmonic factor $\caH_k^f$. Namely, we have the block decomposition
\[
\Gf^{\sigma} \cong \bigoplus_{k=0}^n \left(T_k^\sigma \otimes \Id_{\caH_k^f}\right) \mcomma
\]
where $T_k^\sigma \in \End(V_{n-k})$ denotes the action of $\Gf^\sigma$ on $V_{n-k}$.
\end{prop}

\begin{proof}
By De Bie's harmonic Gaussian identity in \cite[Theorem 4]{DB08}, we have
\[
 \Gf^\sigma\widetilde H_{k,l}^f
   = (\sigma\compi)^k\widetilde H_{k,l}^f \mdot
\]
Meanwhile, by the first-order intertwining relations in Proposition \ref{prop:intertwining}, we check that $\Gf^\sigma X = Y\Gf^\sigma$, $\Gf^\sigma Y = X\Gf^\sigma$, and $\Gf^\sigma H = -H\Gf^\sigma$.
Since $\omega(E_+) = \frac{1}{2}(X-Y+H)$, it follows that $\Gf^\sigma\omega(E_+) = -\omega(E_+)\Gf^\sigma$.
Applying $\Gf^\sigma$ to $\psi_{j,k,l} = (4\omega(E_+))^j\widetilde H_{k,l}^f$, we obtain
\[
 \Gf^\sigma\psi_{j,k,l}
 = (-1)^j(\sigma\compi)^k\psi_{j,k,l}
 = (\sigma\compi)^{2j+k}\psi_{j,k,l} \mdot
\]
Since the eigenvalue $(\sigma\compi)^{2j+k}$ does not depend on the harmonic index $l$, we get the stated block decomposition. Furthermore, since $\omega(K)\psi_{j,k,l} = (2j+k-n)\psi_{j,k,l}$ on the basis $\{\psi_{j,k,l}\}$ of $\Lambda_{2n}$, we obtain the operator exponential formula
\[
 \Gf^\sigma
 = (\sigma\compi)^n
   \exp\!\left(\frac{\sigma\compi\pi}{2}\omega(K)\right)
 = \exp\!\left(\frac{\sigma\compi\pi}{4}
       (\Delta_f-\bm{\grave{x}}^{\,2}+2n)\right) \mcomma
\]
which agrees with the fermionic case of \cite[Theorem 10]{DB08}. This finishes the proof.
\end{proof}

\begin{corl}
For each fixed $k$ and $l$, with respect to the Clifford--Hermite basis $\{\psi_{j,k,l}\}_{j=0}^{n-k}$, the radial operator $T_k^\sigma \in \End(V_{n-k})$ is diagonal with eigenvalues
\[
\lambda_{j,k}^\sigma = (\sigma \compi)^{2j+k} \mcomma \qquad 0 \le j \le n-k \mdot
\]
\end{corl}

\begin{thm}\label{thm:Lambda2n_operatordecom}
Let $\caT \in \End(\caP)$ be a linear operator satisfying the following three conditions:
\begin{enumerate}
    \item[(a)] $\caT$ is $Sp(n)$-equivariant, i.e., $\caT(g \cdot f) = g \cdot \caT(f)$ for all $f \in \caP$ and $g \in Sp(n)$;
    \item[(b)] $\caT$ satisfies the Helmholtz relations with respect to the fermionic Laplacian $\Delta_f$ and the quadratic invariant $\bm{\grave{x}}^2 = \sum_{j=1}^n \grave{x}_{2j-1}\grave{x}_{2j}$:
    $$ \caT \circ \bm{\grave{x}}^2 = -\Delta_f \circ \caT \quad \text{and} \quad \caT \circ \Delta_f = -\bm{\grave{x}}^2 \circ \caT \mcomma $$
    \item[(c)] $\caT^4 = \Id$.
\end{enumerate}
Then, under the decomposition in Proposition \ref{prop:fourier_decomp}, $\caT$ has the unique block form
\[
\caT
\cong
\bigoplus_{k=0}^n
c_k\left(
T_k^+\otimes\Id_{\caH_k^f}
\right) \mcomma
\]
where $c_k \in \{\pm 1, \pm \compi\}$ for each $k \in \{0, \dots, n\}$.
Conversely, every operator of this block form satisfies conditions \textup{(a)}--\textup{(c)}.
\end{thm}

\begin{proof}
Let us put
\[
\widetilde{\caT} \coloneqq \caT \circ (\Gf^{+})^{-1} \mdot
\]
In terms of $X = \frac{1}{2}\bm{\grave{x}}^2$ and $Y = -\frac{1}{2}\Delta_f$, condition (b) means that
$$ \caT \circ X = Y \circ \caT \quad \text{and} \quad \caT \circ Y = X \circ \caT \mdot $$
Using $[X,Y]=H$, we get
$$ \caT \circ H = \caT \circ (XY - YX) = YX\caT - XY\caT = -[X, Y] \circ \caT = -H \circ \caT \mdot $$
Since $\Gf^{+}$ satisfies the exact same relations with $X$, $Y$, and $H$, we see that $\widetilde{\caT}$ commutes with $X$, $Y$, and $H$, and hence $\widetilde{\caT}$ is $\fsl_2(\bbC)$-equivariant.

By condition (a) and Proposition \ref{prop:fourier_decomp}, both $\caT$ and $\Gf^{+}$ commute with the action of $Sp(n)$, so $\widetilde{\caT}$ also commutes with $Sp(n)$. Therefore, $\widetilde{\caT}$ commutes with the joint action of $\fsl_2(\bbC) \times Sp(n)$ on $\caP$. By Lemma \ref{lem:dual_pair}, $\caP$ decomposes into the multiplicity-free direct sum of irreducible $\fsl_2(\bbC) \times Sp(n)$-modules $V_{n-k} \otimes \caH_k^f$. By Schur's lemma, for each $k \in \{0, \dots, n\}$, there exists a scalar $c_k \in \bbC$ such that
$$ \widetilde{\caT}\big|_{V_{n-k} \otimes \caH_k^f} = c_k \Id_{V_{n-k} \otimes \caH_k^f} \mdot $$
Combining this with $\caT = \widetilde{\caT} \circ \Gf^{+}$ and Proposition \ref{prop:fourier_decomp}, we obtain
$$ \caT = \widetilde{\caT} \circ \Gf^{+} \cong \bigoplus_{k=0}^n c_k (T_k^+ \otimes \Id_{\caH_k^f}) \mdot $$

Finally, we consider condition (c). Since $\caT^4 = \Id$ and $(\Gf^{+})^4 = \Id$, we have $(T_k^+)^4 = \Id_{V_{n-k}}$ on each block. Taking the fourth power of the block decomposition of $\caT$, we get $c_k^4 = 1$, i.e., $c_k \in \{\pm 1, \pm \compi\}$ for all $k \in \{0, \dots, n\}$. Conversely, since each summand $V_{n-k}\otimes\caH_k^f$ is invariant under $X$, $Y$, and $Sp(n)$, any such block operator with $c_k \in \{\pm 1, \pm \compi\}$ satisfies conditions (a)--(c). This finishes the proof.
\end{proof}


\section{Hybrid Fourier transform: covariance and rigidity}
\label{sec:hybrid}\label{sec:prelims}\label{sec:rigidity}

\subsection{Orthosymplectic Lie superalgebra and hybrid Fourier transform}
In this section, we follow the conventions of De Bie \cite[Sections 2--4]{DB08} for the superspace $\bbR^{m|2n}$. Let us put $M \coloneqq m-2n$; when $m=0$, we set $\caS(\bbR^0)=\bbC$ and $\Ub^\sigma=\Id$.
We denote the commuting coordinates by $x_1,\ldots,x_m$ and the anticommuting Grassmann coordinates by $\grx_1,\ldots,\grx_{2n}$, and we write $\bm{\breve{x}} \coloneqq (\bmx,\bmgrx)$. Here, the partial derivatives $\partial_{x_r}$ act on the first tensor factor, while the left Grassmann derivatives $\partial_{\grx_a}$ act on the second factor as in Section \ref{sec:grassmann}. We define the polynomial space $\caP$ and the Schwartz--Grassmann space $\caS(\bbR^{m|2n})$ by
\[
 \caP \coloneqq \bbC[x_1,\ldots,x_m]\otimes\Lambda_{2n} \mcomma
\]
\begin{equation}\label{eq:functionspace}
 \caS(\bbR^{m|2n}) \coloneqq \caS(\bbR^m)\otimes\Lambda_{2n} \mdot
\end{equation}
Every superfunction $f \in \caS(\bbR^{m|2n})$ has a unique expansion $f = \sum_I f_I(\bmx)\grx_I$ with coefficients $f_I \in \caS(\bbR^m)$. Since $\Lambda_{2n}$ has dimension $2^{2n}$, $\caS(\bbR^{m|2n})$ is a finite direct sum of copies of $\caS(\bbR^m)$ equipped with its natural Fr\'echet topology.

\begin{defn}\label{defn:euler}
We define the superspace bilinear pairing and the generalized squared radius $R^2$ by
\begin{align}
 \langle\bm{\breve{x}},\bm{\breve{y}}\rangle
 &\coloneqq -\sum_{r=1}^m x_ry_r+
   \frac12\sum_{j=1}^n
   (\grx_{2j-1}\gry_{2j}-\grx_{2j}\gry_{2j-1}) \mcomma
   \label{eq:innerproduct}\\
 R^2 &\coloneqq \langle\bm{\breve{x}},\bm{\breve{x}}\rangle
 = \bmgrx^2-|\bmx|^2 \mcomma
 \qquad \bmgrx^2 = \sum_{j=1}^n\grx_{2j-1}\grx_{2j} \mdot
 \label{eq:generalized_radius}
\end{align}
We define the scalar super-Laplacian $\Delta$ and the super-Euler operator $\bbE$ by
\[
 \Delta \coloneqq \Delta_b+\Delta_f \mcomma
 \qquad \Delta_b \coloneqq -\sum_{r=1}^m\partial_{x_r}^2 \mcomma
 \qquad \Delta_f \coloneqq 4\sum_{j=1}^n
                   \partial_{\grx_{2j-1}}\partial_{\grx_{2j}} \mcomma
\]
\[
 \bbE \coloneqq \bbE_b+\bbE_f \mcomma
 \qquad \bbE_b \coloneqq \sum_{r=1}^m x_r\partial_{x_r} \mcomma
 \qquad \bbE_f \coloneqq \sum_{a=1}^{2n}\grx_a\partial_{\grx_a} \mdot
\]
\end{defn}

Note that the pairing $\langle\bm{\breve{x}},\bm{\breve{y}}\rangle$ is an even supersymmetric bilinear form: since the odd coordinates anticommute, we have $\langle\bm{\breve{x}},\bm{\breve{y}}\rangle = \langle\bm{\breve{y}},\bm{\breve{x}}\rangle$.
If we set
\[
 E_+ \coloneqq R^2/2 \mcomma \qquad E_- \coloneqq -\Delta/2 \mcomma \qquad
 H \coloneqq \bbE+M/2 \mcomma
\]
then by direct calculation we get the $\fsl_2$-commutation relations
\begin{equation}\label{eq:sl2}
 [E_+,E_-] = H \mcomma \qquad [H,E_+] = 2E_+ \mcomma \qquad [H,E_-] = -2E_- \mdot
\end{equation}
When $(m,n)\ne(0,0)$, these operators span a Lie algebra isomorphic to $\fsl_2(\bbC)$. Note that here $E_\pm$ denote the coordinate radial operators on $\caS(\bbR^{m|2n})$, rather than the rotated fermionic basis elements in Section \ref{sec:grassmann}.

\begin{defn}\label{defn:osp}
Write $X = (X_1,\ldots,X_{m+2n}) \coloneqq (x_1,\ldots,x_m,\grx_1,\ldots,\grx_{2n})$, and denote the parity of the index $j$ by $|j|=0$ for $1\le j\le m$ and $|j|=1$ for $m+1\le j\le m+2n$. The matrix of the superspace pairing (\ref{eq:innerproduct}) is given by
\begin{equation}\label{eq:orthosymplectic_metric}
 g \coloneqq \operatorname{diag}(-I_m,J) \mcomma \qquad
 J \coloneqq \frac12\operatorname{diag}
 \left(\begin{pmatrix}0&1\\-1&0\end{pmatrix},\ldots,
       \begin{pmatrix}0&1\\-1&0\end{pmatrix}\right) \mdot
\end{equation}
The \textbf{orthosymplectic Lie superalgebra} $\fosp(m|2n)$ is the Lie superalgebra preserving this even bilinear form. Writing $(g^{jk})=g^{-1}$ and $\partial_{X^j} \coloneqq \sum_k g^{jk}\partial_{X_k}$, we obtain the coordinate differential-operator realization of $\fosp(m|2n)$ spanned by
\begin{equation}\label{eq:osprealization}
 L_{jk} \coloneqq X_j\partial_{X^k}
       -(-1)^{|j||k|}X_k\partial_{X^j} \mcomma
       \qquad 1\le j\le k\le m+2n \mdot
\end{equation}
In particular, we have $L_{jj}=0$ for $1\le j\le m$ and $L_{jj}=2X_j\partial_{X^j}$ for $m+1\le j\le m+2n$.
\end{defn}

In this coordinate realization (cf. \cite{Z08, C13}), the generators $L_{jk}$ commute with $R^2$, $\Delta$, and $\bbE$, forming the coordinate super dual pair $(\fosp(m|2n), \fsl_2)$ with even part $\fso(m)\oplus\fsp(2n)$.

Combining the Lebesgue integral on $\bbR^m$ with the normalized Berezin integral (\ref{eq:fermionic-berezin}), we define the Berezin--Lebesgue integral on $\caS(\bbR^{m|2n})$ by
\begin{equation}\label{eq:BerezinOnly}
 \int_{\bbR^{m|2n}} f
 \coloneqq \int_{\bbR^m}d\bmx\int_{B,\bmgrx}f
 = \pi^{-n}\int_{\bbR^m}f_{\{1,\ldots,2n\}}(\bmx)\,d\bmx \mdot
\end{equation}
Note that the $2^{2n}$ ordered Grassmann monomials
\begin{equation}\label{eq:monomial_basis}
 \{\grx_1^{\epsilon_1}\cdots\grx_{2n}^{\epsilon_{2n}} :\,
       \epsilon\in\{0,1\}^{2n}\}
\end{equation}
allow us to identify $\Lambda_{2n}$ as a vector space with the $2n$-qubit space $(\bbC^2)^{\otimes2n}$.

\begin{defn}\label{defn:hybrid-transform}
Following \cite[Section 4]{DB08}, for $\sigma \in \{+1,-1\}$, we define the \textbf{hybrid Fourier transform} (or superspace Fourier transform) on $\caS(\bbR^{m|2n})$ by
\begin{equation}\label{eq:hybrid-definition}
 \Fhyb^\sigma f(\bmy,\bmgry)
 \coloneqq (2\pi)^{-M/2}\int_{\bbR^m}d\bmx\int_{B,\bmgrx}
   e^{-\sigma\compi\langle\bm{\breve{x}},\bm{\breve{y}}\rangle}
   f(\bmx,\bmgrx) \mdot
\end{equation}
\end{defn}

Since the exponential kernel $e^{-\sigma\compi\langle\bm{\breve{x}},\bm{\breve{y}}\rangle}$ factors as the product of the bosonic kernel $e^{\sigma\compi\bmx\cdot\bmy}$ and the fermionic kernel $e^{-\sigma\compi\langle\bmgrx,\bmgry\rangle}$, and $(2\pi)^{-M/2} = (2\pi)^{-m/2}(2\pi)^n$, we see immediately that $\Fhyb^\sigma$ factors as the tensor product
\[
 \Fhyb^\sigma = \Ub^\sigma\otimes\Gf^\sigma \mcomma
 \qquad (\Fhyb^\sigma)^{-1} = \Fhyb^{-\sigma} \mdot
\]
In particular, $\Fhyb^\sigma$ is a continuous linear automorphism of $\caS(\bbR^{m|2n})$.

\subsection{Covariance and Howe decomposition of hybrid Fourier transform}
\begin{prop}\emph{\cite[Section 4, Lemma 3]{DB08}}
\label{prop:generalFT}
For $\sigma\in\{+1,-1\}$, $1\le r\le m$, and $1\le j\le n$, we have the first-order intertwining relations
\begin{align*}
\Fhyb^{\sigma}\partial_{x_r}&=-\sigma \compi\, y_r\Fhyb^{\sigma}\mcomma
&\Fhyb^{\sigma}x_r&=-\sigma \compi\,\partial_{y_r}\Fhyb^{\sigma}\mcomma\\
\Fhyb^{\sigma}\partial_{\grx_{2j}}&=-\sigma \frac{\compi}{2}\,\gry_{2j-1}\Fhyb^{\sigma}\mcomma
&\Fhyb^{\sigma}\partial_{\grx_{2j-1}}&=\sigma \frac{\compi}{2}\,\gry_{2j}\Fhyb^{\sigma}\mcomma\\
\Fhyb^{\sigma}\grx_{2j}&=2\sigma\compi\,\partial_{\gry_{2j-1}}\Fhyb^{\sigma}\mcomma
&\Fhyb^{\sigma}\grx_{2j-1}&=-2\sigma\compi\,\partial_{\gry_{2j}}\Fhyb^{\sigma}\mdot
\end{align*}
\end{prop}

From Proposition \ref{prop:generalFT}, we deduce the following second-order covariance relations for $\Fhyb^{\sigma}$:
\begin{align}
 \Fhyb^\sigma R^2 &= -\Delta\Fhyb^\sigma \mcomma &
 \Fhyb^\sigma\Delta &= -R^2\Fhyb^\sigma \mcomma
 \label{eq:radial-fourier}\\
 \Fhyb^\sigma(\bbE+M/2) &= -(\bbE+M/2)\Fhyb^\sigma \mcomma &
 \Fhyb^\sigma L_{jk} &= L_{jk}\Fhyb^\sigma \mdot
 \label{eq:coordinate-fourier}
\end{align}
Namely, conjugation by $\Fhyb^{\sigma}$ exchanges $E_+$ and $E_-$, sends $H$ to $-H$, and commutes with the coordinate action of $\fosp(m|2n)$.

Indeed, using the first-order intertwining relations in Proposition \ref{prop:generalFT}, we check directly that conjugation by $\Fhyb^\sigma$ sends $R^2$ to $-\Delta$, $\Delta$ to $-R^2$, $\bbE_b$ to $-\bbE_b-m$, and $\bbE_f$ to $2n-\bbE_f$. Moreover, the even generators of $\fosp(m|2n)$ commute with $\Fhyb^\sigma$ by the $O(m)$-equivariance of $\Ub^\sigma$ and the $Sp(n)$-equivariance of $\Gf^\sigma$. For the odd generators of $\fosp(m|2n)$, let us put $a=2j-1$ and $b=2j$. By (\ref{eq:orthosymplectic_metric}) and (\ref{eq:osprealization}), we have
\[
 L_{r,m+a} = -2x_r\partial_{\grx_b}+\grx_a\partial_{x_r} \mcomma \qquad
 L_{r,m+b} = 2x_r\partial_{\grx_a}+\grx_b\partial_{x_r} \mdot
\]
Applying Proposition \ref{prop:generalFT}, we see that conjugation by $\Fhyb^\sigma$ sends these two operators to
\[
 \gry_a\partial_{y_r}-2y_r\partial_{\gry_b} \mcomma \qquad
 \gry_b\partial_{y_r}+2y_r\partial_{\gry_a} \mcomma
\]
which are precisely $L_{r,m+a}$ and $L_{r,m+b}$ in the target variables.

The Howe decompositions related to the superspace Fourier transform were studied in \cite{DB08, C13, Z08}. Note that on a mixed superspace, when the even superdimension $M = m-2n$ lies in $-2\bbN$, the full coordinate Fischer decomposition can fail to be a direct sum; however, we can always combine the bosonic Howe decomposition (\ref{eq:bosonic-howe-decomposition}) with the fermionic Lefschetz decomposition of Lemma \ref{lem:dual_pair} as follows.

\begin{prop}
\label{prop:hybrid-block-decomposition}
We have the $Sp(n)$-equivariant decomposition
\begin{equation}
 \caS(\bbR^{m|2n})
 \cong \bigoplus_{k=0}^n
 \left(\caS(\bbR^m)\otimes V_{n-k}\right)
 \otimes\caH_k^f \mcomma
 \label{eq:hybrid-space-blocks}
\end{equation}
under which the hybrid Fourier transform $\Fhyb^\sigma$ decomposes as
\begin{equation}
 \Fhyb^\sigma \cong \bigoplus_{k=0}^n
 \left[
  \left(\Ub^\sigma\otimes T_k^\sigma\right)
  \otimes\Id_{\caH_k^f}
 \right] \mdot
 \label{eq:hybrid-fourier-blocks}
\end{equation}
For $m\geq1$, let $\caS_{\rm fin}(\bbR^{m|2n}) \coloneqq \caS_{\rm fin}(\bbR^m)\otimes\Lambda_{2n}$. Combining the decompositions of both tensor factors, we get the algebraic direct sum decomposition
\begin{equation}
 \caS_{\rm fin}(\bbR^{m|2n})
 \cong
 \bigoplus_{p\in\mathcal K_m}\bigoplus_{k=0}^n
 \left(D_{p+m/2}\otimes V_{n-k}\right)
 \otimes\left(\caH_p^b\otimes\caH_k^f\right)
 \label{eq:hybrid-double-blocks}
\end{equation}
under the action of $(\fsl_2^b\oplus\fsl_2^f) \times (O(m)\times Sp(n))$, and we have
\begin{equation}
 \Fhyb^\sigma
 \cong
 \bigoplus_{p\in\mathcal K_m}\bigoplus_{k=0}^n
 \left[
 \left(
  B_p^\sigma\otimes T_k^\sigma
 \right)
 \otimes\Id_{\caH_p^b\otimes\caH_k^f}
 \right] \mdot
 \label{eq:hybrid-double-fourier-blocks}
\end{equation}
In particular, if $h_{\bm\alpha}$ is a multivariate Hermite function on $\bbR^m$ and $\psi_{r,k,\ell}$ is the Clifford--Hermite basis element from Lemma \ref{lem:clifford_hermite_basis}, then we have
\begin{equation}
 \Fhyb^\sigma
 (h_{\bm\alpha}\otimes\psi_{r,k,\ell})
 = (\sigma\compi)^{|\bm\alpha|+2r+k}
  h_{\bm\alpha}\otimes\psi_{r,k,\ell} \mdot
 \label{eq:hybrid-spectrum}
\end{equation}
\end{prop}
\begin{proof}
By tensoring the fermionic Lefschetz decomposition of Lemma \ref{lem:dual_pair} with $\caS(\bbR^m)$ and using $\Fhyb^\sigma = \Ub^\sigma\otimes\Gf^\sigma$ together with Proposition \ref{prop:fourier_decomp}, we get (\ref{eq:hybrid-space-blocks}) and (\ref{eq:hybrid-fourier-blocks}). Similarly, by tensoring Lemma \ref{lem:dual_pair} with the bosonic Howe decomposition (\ref{eq:bosonic-howe-decomposition}) on $\caS_{\rm fin}(\bbR^m)$ and regrouping the tensor factors, we obtain (\ref{eq:hybrid-double-blocks}) and (\ref{eq:hybrid-double-fourier-blocks}). Finally, combining the bosonic eigenvalue formula (\ref{eq:bosonic-fourier-eigenvalues}) with the fermionic eigenvalue formula in Proposition \ref{prop:fourier_decomp}, we get (\ref{eq:hybrid-spectrum}). This finishes the proof.
\end{proof}

\begin{remark}
Note that (\ref{eq:hybrid-space-blocks}) holds for all $m,n\ge0$, while (\ref{eq:hybrid-double-blocks}) holds on the dense subspace of Hermite-finite vectors $\caS_{\rm fin}(\bbR^{m|2n})$ for $m\ge1$. Since the odd generators of the coordinate Lie superalgebra $\fosp(m|2n)$ mix the bosonic and fermionic factors, (\ref{eq:hybrid-double-blocks}) is a decomposition with respect to the even subgroup $O(m)\times Sp(n)$ rather than the full coordinate $\fosp(m|2n)$-action.
\end{remark}

\subsection{The quadratic oscillator algebra and its commutants}
Now we assume $m,n\ge1$. In the rest of this section, we identify the source and target variables and write $(\bmx,\bmgrx)$ for the coordinates on $\bbR^{m|2n}$. Let us denote the spaces of first-order bosonic and fermionic canonical generators by
\[
 E_b \coloneqq \spanC\{x_r,\partial_{x_r} :\, 1\le r\le m\} \mcomma \qquad
 E_f \coloneqq \spanC\{\grx_a,\partial_{\grx_a} :\, 1\le a\le2n\} \mdot
\]
The Weyl commutation relations define a nondegenerate alternating bilinear form on $E_b$, while the Clifford anticommutation relations define a nondegenerate symmetric bilinear form on $E_f$. We map quadratic tensors to differential operators on $\caS(\bbR^{m|2n})$ via the symmetrized and antisymmetrized products
\[
 u\odot v \longmapsto \tfrac12(uv+vu) \mcomma \qquad
 a\wedge b \longmapsto \tfrac12(ab-ba) \mcomma \qquad
 u\otimes a \longmapsto ua \mdot
\]
We denote the even, odd (mixed), and total quadratic operator spaces by
\[
 \frg_0 \coloneqq S^2E_b\oplus\bigwedge^2E_f \mcomma
 \qquad \frg_1 \coloneqq E_b\otimes E_f \mcomma
 \qquad \frg \coloneqq \frg_0\oplus\frg_1 \mdot
\]
In particular, $\frg_0$ contains the centered operators $x_r\partial_{x_s}+\tfrac12\delta_{rs}$ and $\grx_a\partial_{\grx_b}-\tfrac12\delta_{ab}$.
Let us define the bosonic, fermionic, and total parity operators on $\caS(\bbR^{m|2n})$ by
\[
 (\Pb f)(\bmx,\bmgrx) \coloneqq f(-\bmx,\bmgrx) \mcomma \qquad
 (\Pf f)(\bmx,\bmgrx) \coloneqq f(\bmx,-\bmgrx) \mcomma \qquad \Pbf \coloneqq \Pb\Pf \mdot
\]
For $s,t\in\{\pm1\}$, we define the parity subspaces and projections by
\[
 \caF_t \coloneqq \ker(\Pf-t\Id)|_{\Lambda_{2n}} \mcomma \qquad
 \Pi_t^f \coloneqq \tfrac12(\Id+t\Pf) \mcomma \qquad
 \Pi_{s,t} \coloneqq \tfrac14(\Id+s\Pb)(\Id+t\Pf) \mdot
\]
Since $m,n\ge1$, all four joint parity sectors $\Pi_{s,t}\caS(\bbR^{m|2n})$ are nonzero. In the following lemma, we compute the ordinary (ungraded) commutants of $\frg_0$ and $\frg$ inside the algebra of continuous linear endomorphisms of $\caS(\bbR^{m|2n})$.

\begin{lemma} 
\label{lem:quadratic-oscillator}\label{lem:full-even-commutant}
The space of quadratic operators $\frg = \frg_0 \oplus \frg_1$ is a complex Lie superalgebra under the supercommutator, and we have the isomorphism
\[
 \frg \cong \fosp(4n|2m) \mcomma \quad \text{with even part }
 \frg_0 \cong \fso(4n)\oplus\fsp(2m) \mdot
\]
On the Schwartz--Grassmann space $\caS(\bbR^{m|2n})$, the continuous ordinary commutants of $\frg_0$ and $\frg$ are given by
\begin{align}
 \End^{\rm cont}_{\frg_0}\bigl(\caS(\bbR^{m|2n})\bigr)
 &= \bbC[\Pb,\Pf]
   = \spanC\{\Id,\Pb,\Pf,\Pbf\} \mcomma \label{eq:even-commutant}\\
 \End^{\rm cont}_{\frg}\bigl(\caS(\bbR^{m|2n})\bigr)
 &= \bbC[\Pbf]
   = \spanC\{\Id,\Pbf\} \mdot \label{eq:full-commutant}
\end{align}
The same formulas hold for the algebraic commutants on the Hermite-finite subspace $\caS_{\rm fin}(\bbR^{m|2n})$.
Furthermore, under the action of $\frg$, the two total-parity subspaces of $\caS_{\rm fin}(\bbR^{m|2n})$ are inequivalent irreducible $\frg$-modules, and their closures in $\caS(\bbR^{m|2n})$ are the two total-parity subspaces of $\caS(\bbR^{m|2n})$.
\end{lemma}
\begin{proof}
The even summands $S^2E_b \cong \fsp(2m)$ and $\bigwedge^2E_f \cong \fso(4n)$ are the standard oscillator and spin representations. To check the supercommutator of two mixed quadratic elements in $\frg_1 = E_b\otimes E_f$, let us write $[u,v]=\omega(u,v)\Id$ for $u,v\in E_b$ and $\{a,b\}=\eta(a,b)\Id$ for $a,b\in E_f$. Then we have
\[
 \{ua,vb\}
 = \eta(a,b)\tfrac12(uv+vu)
   +\omega(u,v)\tfrac12(ab-ba) \in \frg_0 \mdot
\]
Together with the natural action of $\frg_0$ on $\frg_1$, this gives the defining supercommutation relations of $\fosp(4n|2m)$. By writing operators in terms of normally ordered Weyl--Clifford monomials, we see that this representation of $\fosp(4n|2m)$ is faithful (cf. \cite{Nis90} and \cite[Section 5.6]{C13}).

Now let $A$ be a continuous linear endomorphism of $\caS(\bbR^{m|2n})$ that commutes with $\frg_0$. In particular, $A$ commutes with each bosonic number operator and each fermionic occupation number operator:
\[
 N_r \coloneqq \tfrac12(x_r^2-\partial_{x_r}^2-1) \quad (1\le r\le m) \mcomma \qquad
 n_a \coloneqq \grx_a\partial_{\grx_a} \quad (1\le a\le 2n) \mcomma
\]
because $N_r+\tfrac12\Id$ and $n_a - \tfrac12\Id$ lie in $\frg_0$. In $\caS(\bbR^{m|2n})$, the joint eigenspaces of $\{N_r, n_a\}$ are the one-dimensional lines spanned by the Hermite--Grassmann basis elements $h_{\bm k}\otimes\grx_I$. Therefore, $A$ acts diagonally on this basis:
\[
 A(h_{\bm k}\otimes\grx_I)
 = a_{\bm k,I}\,h_{\bm k}\otimes\grx_I \mdot
\]
Since the even quadratic operators in $S^2E_b$ (respectively, $\bigwedge^2E_f$) act irreducibly on each parity sector of $\caS_{\rm fin}(\bbR^m)$ (respectively, $\Lambda_{2n}$), the eigenvalue $a_{\bm k,I}$ depends only on the parities $(|\bm k|\bmod2,\, |I|\bmod2)$. Thus we have $A = \sum_{s,t\in\{\pm1\}} a_{s,t}\Pi_{s,t}$ on $\caS_{\rm fin}(\bbR^{m|2n})$.

If $A$ also commutes with the odd subspace $\frg_1$, then the mixed operators in $\frg_1$ shift both $|\bm k|$ and $|I|$ by $\pm 1$, which forces $a_{s,t} = a_{-s,-t}$. Thus only two independent coefficients remain, so $A \in \spanC\{\Id,\Pbf\}$ on $\caS_{\rm fin}(\bbR^{m|2n})$. Conversely, it is clear that $\Pb$ and $\Pf$ commute with $\frg_0$, and $\Pbf$ commutes with all of $\frg$. Since $\caS_{\rm fin}(\bbR^{m|2n})$ is dense in $\caS(\bbR^{m|2n})$ and $A$ is continuous, both commutant formulas (\ref{eq:even-commutant}) and (\ref{eq:full-commutant}) extend to $\caS(\bbR^{m|2n})$.

Finally, let us check the irreducibility of the two total-parity subspaces of $\caS_{\rm fin}(\bbR^{m|2n})$. Any nonzero $\frg$-invariant subspace $W \subseteq \caS_{\rm fin}(\bbR^{m|2n})$ contains a nonzero finite linear combination of basis vectors $h_{\bm k}\otimes\grx_I$. Using polynomial interpolation in the commuting operators $\{N_r, n_a\}$, we see that $W$ contains at least one basis vector $h_{\bm k}\otimes\grx_I$. Applying the even and mixed quadratic operators in $\frg$, we can reach every basis vector of the same total parity. Moreover, since the joint spectra of $\{N_r, n_a\}$ on the two total-parity subspaces are disjoint, the two $\frg$-modules are inequivalent. This finishes the proof.
\end{proof}

\subsection{Rigidity of hybrid Fourier transform}

Let us write
$$\kappa \coloneqq \operatorname{Conj}(\Fhyb^+)$$
for the conjugation map $Q \longmapsto \Fhyb^+ Q (\Fhyb^+)^{-1}$. Its action on the first-order generators is given by Proposition \ref{prop:generalFT} with $\sigma=+1$ (after identifying the target variables with the source variables), and its action on higher-order operators is obtained multiplicatively. Consider the coordinate family of pure and mixed quadratic operators
\begin{equation}\label{eq:quadratic-family}
\mathcal Q \coloneqq \{x_rx_s,\ \partial_{x_r}\partial_{x_s},\
 \grx_a\grx_b,\ \partial_{\grx_a}\partial_{\grx_b},\
 x_r\grx_a,\ \partial_{x_r}\partial_{\grx_a} :\,
 1\le r,s\le m,\quad 1\le a,b\le2n\} \mdot
\end{equation}
Using $\kappa$, we can write the coordinate quadratic intertwining relations for an operator $T$ in the compact form $TQ = \kappa(Q)T$. For example, on the mixed quadratic generators, $TQ = \kappa(Q)T$ means
\[
 T(x_r\grx_{2j}) = 2\partial_{x_r}\partial_{\grx_{2j-1}}T \mcomma
 \qquad
 T(\partial_{x_r}\partial_{\grx_{2j}})
       = -\tfrac12x_r\grx_{2j-1}T \mdot
\]
For any operator $Q$ preserving $\caS(\bbR^{m|2n})$, multiplying on the left by $(\Fhyb^+)^{-1}$ gives the equivalence
\begin{equation}\label{eq:basic-commutant}
 TQ = \kappa(Q)T
 \quad\Longleftrightarrow\quad
 [(\Fhyb^+)^{-1}T,\, Q] = 0 \mdot
\end{equation}

\begin{thm}
\label{thm:generalFT}\label{thm:one-mixed}\label{lem:q0-generation}
Let $T$ be a continuous complex-linear endomorphism of $\caS(\bbR^{m|2n})$.
\begin{enumerate}[(1)]
\item For any $m,n\ge0$, the first-order covariance relations $TZ=\kappa(Z)T$ hold for all canonical generators $Z \in E_b \oplus E_f$ if and only if $T=c\Fhyb^+$ for some $c\in\bbC$.
\item Let $m,n\ge1$ and fix any nonzero mixed quadratic operator $Z\in\frg_1$. Then the following four conditions are equivalent:
\begin{enumerate}[(a)]
\item $TQ=\kappa(Q)T$ for every $Q$ in the coordinate family $\mathcal Q$ of (\ref{eq:quadratic-family});
\item $TQ=\kappa(Q)T$ for every $Q\in\frg$;
\item $TQ=\kappa(Q)T$ for every $Q\in\frg_0$, and $TZ=\kappa(Z)T$;
\item $T=\alpha\Fhyb^++\beta\Fhyb^-$ for unique scalars $\alpha,\beta\in\bbC$.
\end{enumerate}
In particular, under full even quadratic covariance, imposing the intertwining relation for any single nonzero mixed quadratic operator $Z\in\frg_1$ implies that all quadratic intertwining relations hold.
\end{enumerate}
\end{thm}
\begin{proof}
Let us put $A \coloneqq (\Fhyb^+)^{-1}T$. By (\ref{eq:basic-commutant}), $T$ intertwines an operator $Q$ via $\kappa$ if and only if $[A,Q]=0$.

\textbf{(1)} As shown in the proof of Theorem \ref{thm:fermionicFT}(1), the fermionic first-order generators in $E_f$ generate the full matrix algebra $\End(\Lambda_{2n})$. Thus, if $A$ commutes with $E_f$, we have $A = B\otimes\Id_{\Lambda_{2n}}$ for some continuous operator $B$ on $\caS(\bbR^m)$. Since $A$ also commutes with the bosonic generators in $E_b$, the operator $B$ commutes with $x_r$ and $\partial_{x_r}$ for all $1\le r\le m$. In particular, $B$ preserves the common kernel of $\partial_{x_r}+x_r$ (the one-dimensional Gaussian line $\bbC e^{-|\bmx|^2/2}$) and commutes with the creation operators $x_r-\partial_{x_r}$. Hence $B$ acts as the same scalar $c\in\bbC$ on every Hermite function. By the density of Hermite functions in $\caS(\bbR^m)$ and the continuity of $B$, we get $B = c\Id$, so $T = c\Fhyb^+$. The boundary cases $m=0$ or $n=0$ follow immediately, and the converse holds by Proposition \ref{prop:generalFT}.

\textbf{(2)} First, let us show that the coordinate family $\mathcal Q$ in (\ref{eq:quadratic-family}) generates the entire quadratic Lie superalgebra $\frg$ under commutators and supercommutators. Indeed, the commutators of $x_rx_s$ and $\partial_{x_r}\partial_{x_s}$ generate all centered transfer operators $x_r\partial_{x_s}+\tfrac12\delta_{rs}$, so the bosonic members of $\mathcal Q$ generate $S^2E_b$. Taking the commutators of $S^2E_b$ with the mixed operators $x_r\grx_a$ and $\partial_{x_r}\partial_{\grx_a}$ in $\mathcal Q$, we get all of $\frg_1 = E_b\otimes E_f$; for instance, we have
\[
 [\partial_{x_r}^2,\, x_r\grx_a] = 2\partial_{x_r}\grx_a \mcomma
 \qquad
 [x_r^2,\, \partial_{x_r}\partial_{\grx_a}]
     = -2x_r\partial_{\grx_a} \mdot
\]
Next, for any fixed bosonic index $r$, we calculate the supercommutator of $x_r\grx_a$ and $\partial_{x_r}\partial_{\grx_b}$:
\[
 \{x_r\grx_a,\, \partial_{x_r}\partial_{\grx_b}\}
 = \delta_{ab}(x_r\partial_{x_r}+\tfrac12)
  - (\grx_a\partial_{\grx_b}-\tfrac12\delta_{ab}) \mdot
\]
Since $x_r\partial_{x_r}+\tfrac12 \in S^2E_b$ is already generated, this gives every centered fermionic operator $\grx_a\partial_{\grx_b}-\tfrac12\delta_{ab}$ (even when $n=1$). Together with $\grx_a\grx_b$ and $\partial_{\grx_a}\partial_{\grx_b}$, these span $\bigwedge^2E_f$. Thus $\mathcal Q$ generates $\frg$ as a Lie superalgebra. Since $[A,Q_1]=0$ and $[A,Q_2]=0$ imply $[A, Q_1Q_2]=0$, condition (a) implies that $A$ commutes with all of $\frg$, which gives (b).

Clearly (b) implies (c). Now assume that condition (c) holds, so $[A,Q]=0$ for all $Q\in\frg_0$ and $[A,Z]=0$. Under the adjoint action of $\frg_0 \cong \fsp(2m)\oplus\fso(4n)$, the odd space $\frg_1 = E_b\otimes E_f$ is the external tensor product of the standard representation $E_b$ of $\fsp(2m)$ and the vector representation $E_f$ of $\fso(4n)$. For all $m,n\ge1$, both $E_b$ and $E_f$ are irreducible modules (including the case $n=1$ for $\fso_4(\bbC)$), so $\frg_1$ is an irreducible $\frg_0$-module. Therefore, any nonzero element $Z\in\frg_1$ generates $\frg_1$ under repeated commutators with $\frg_0$. Since $[A,Q]=0$ and $[A,Z]=0$ imply $[A,[Q,Z]]=0$, we see by induction that $A$ commutes with the entire odd space $\frg_1$. Hence $A \in \End_{\frg}^{\rm cont}\bigl(\caS(\bbR^{m|2n})\bigr)$, so (c) implies (b) (and (a)).

Finally, by Lemma \ref{lem:quadratic-oscillator}, $A \in \End_{\frg}^{\rm cont}\bigl(\caS(\bbR^{m|2n})\bigr)$ if and only if $A = \alpha\Id + \beta\Pbf$ for unique scalars $\alpha,\beta\in\bbC$. Since
\[
 \Pbf\Fhyb^+
 = (\Pb\Ub^+)\otimes(\Pf\Gf^+)
 = \Ub^-\otimes\Gf^- = \Fhyb^- \mcomma
\]
we see that $T = \Fhyb^+ A = \alpha\Fhyb^+ + \beta\Fhyb^-$, which proves the equivalence with (d). This finishes the proof.
\end{proof}

\begin{corl}\label{corl:four-sign}
For $m,n\ge1$, an operator $T \in \End^{\rm cont}\bigl(\caS(\bbR^{m|2n})\bigr)$ satisfies the full even quadratic covariance relations $TQ = \kappa(Q)T$ ($Q\in\frg_0$) if and only if
\[
 T = (a\Id+b\Pb+c\Pf+d\Pbf)\Fhyb^+ \mcomma
 \qquad a,b,c,d\in\bbC \mdot
\]
Equivalently, the space of solutions is four-dimensional and is spanned by the four factorwise transforms $\Ub^\varepsilon\otimes\Gf^\delta$ with $\varepsilon,\delta\in\{+1,-1\}$.
Furthermore, when $n\ge2$, covariance under the pure bosonic and pure fermionic quadratics $\{x_rx_s, \partial_{x_r}\partial_{x_s}, \grx_a\grx_b, \partial_{\grx_a}\partial_{\grx_b}\}$ already implies full even covariance, whereas for $n=1$ it does not.
\end{corl}
\begin{proof}
The first assertion follows directly from (\ref{eq:basic-commutant}) and the commutant formula (\ref{eq:even-commutant}) in Lemma \ref{lem:quadratic-oscillator}. The second assertion follows from Lemma \ref{lem:pure-quadratics-generate}.
\end{proof}

\begin{remark}
When $n=1$, the full even quadratic hypothesis in Theorem \ref{thm:generalFT}(2)(c) cannot be weakened to pure quadratics. Indeed, let us write $c_a = \partial_{\grx_a}$ and $c_a^\dagger = \grx_a$. When $n=1$, the Lie algebra generated by the pure fermionic quadratics is $\fl \coloneqq \langle c_1^\dagger c_2^\dagger,\, c_2c_1\rangle \cong \fsl_2(\bbC)$, which acts trivially on $\Lambda_2^{\rm odd}$. If we take the mixed operator $Z = x_1c_1^\dagger$ and the nilpotent operator $N = c_1^\dagger c_2$, we have $N\ne0$, $N^2=0$, $[N,\fl]=0$, and $Nc_1^\dagger = c_1^\dagger N = 0$. Thus $N$ commutes with $\fsp(2m)\oplus\fl$ and with $Z$, but $N \notin \bbC[\Pbf]$. Consequently, $T = \Fhyb^+(\Id+N)$ intertwines $\fsp(2m)\oplus\fl$ and $Z$, but does not lie in $\spanC\{\Fhyb^+,\Fhyb^-\}$.
\end{remark}

\section{Unitary normalization and quantum applications}\label{sec:gate}\label{sec:physical-applications}

In this section, we equip the Grassmann algebra $\Lambda_{2n}$ with the occupation number inner product, in which the ordered monomials $\grx_I$ form an orthonormal basis, and we denote by
\[
\caH_{m|2n} \coloneqq L^2(\bbR^m)\,\widehat\otimes\,\Lambda_{2n}
\]
the Hilbert space completion of $\caS(\bbR^{m|2n})$. Throughout this section, we take $\sigma\in\{+1,-1\}$, and all operator identities involving unbounded operators hold on the common invariant domain $\caS(\bbR^{m|2n})=\caS(\bbR^m)\otimes\Lambda_{2n}$.

\subsection{Normalization in the occupation number basis}\label{subsec:hybrid-realization}

Recall that $\bbE_f=\sum_{a=1}^{2n}\grx_a\partial_{\grx_a}$ is the fermionic Euler operator from Definition \ref{defn:euler}, and $P_f=(-1)^{\bbE_f}$ is the fermionic parity operator. Let us define the degree-scaling operator $\Dn$ on $\Lambda_{2n}$ by
\begin{equation}\label{eq:Dn}
\Dn \coloneqq 2^{-\bbE_f/2} \mcomma \qquad
\Dn\,\grx_I = 2^{-|I|/2}\grx_I \mcomma
\end{equation}
and define the unitarily normalized fermionic Fourier transform by
\begin{equation}\label{eq:Uf}
\Uf^{\sigma} \coloneqq \Dn^{-1}\Gf^{\sigma}\Dn \mdot
\end{equation}
Note that conjugating $\Gf^\sigma$ by $\Dn$ is equivalent to passing from the weighted inner product
\[
 \langle f,g\rangle_D
 \coloneqq \langle\Dn^{-1}f,\Dn^{-1}g\rangle
 = \sum_I 2^{|I|}\overline{f_I}g_I
\]
to the standard occupation number inner product $\langle f,g\rangle = \sum_I \overline{f_I}g_I$. Up to an overall positive normalization and the choice of which argument is complex linear, $\langle\cdot,\cdot\rangle_D$ is the fermionic inner product of \cite[Definition 2.10]{CD11}; see also \cite{CDS11}.

\begin{prop}\label{prop:onepair}\label{thm:gate-factor}
For $n=1$, let us identify the two-qubit occupation basis vector $|\epsilon_1\epsilon_2\rangle$ with the Grassmann monomial $\grx_1^{\epsilon_1}\grx_2^{\epsilon_2}$. With respect to the ordered basis $(|00\rangle,|01\rangle,|10\rangle,|11\rangle)$, the matrix of $\Uf^{\sigma}$ is given by
\[
G_{\sigma} =
\begin{pmatrix}
0&0&0&1\\
0&\sigma\compi&0&0\\
0&0&\sigma\compi&0\\
1&0&0&0
\end{pmatrix} \mdot
\]
\end{prop}
\begin{proof}
For $n=1$, by expanding the Berezin integral in (\ref{defn:fermionicFT}) directly, we have
\[
\Gf^{\sigma}(1) = \tfrac12\gry_1\gry_2 \mcomma \qquad
\Gf^{\sigma}(\grx_a) = \sigma\compi\gry_a \quad (a=1,2) \mcomma \qquad
\Gf^{\sigma}(\grx_1\grx_2) = 2 \mdot
\]
Conjugating this matrix by $D_1 = \operatorname{diag}(1,2^{-1/2},2^{-1/2},2^{-1})$, we immediately get the unitary matrix $G_\sigma$, whose square acts as $+\Id$ on the even subspace $\vectorspan\{|00\rangle, |11\rangle\}$ and as $-\Id$ on the odd subspace $\vectorspan\{|01\rangle, |10\rangle\}$. For general $n \geq 1$, since the symplectic pairing (\ref{eq:fermionic-pairing}), the Berezin integral, and the scaling operator $\Dn$ all factor across the $n$ symplectic pairs $(\grx_{2j-1}, \grx_{2j})$, we see that $\Uf^\sigma$ factors as the $n$-fold tensor power of $G_\sigma$. This finishes the proof.
\end{proof}

Under the tensor product decomposition into $n$ symplectic pairs
\[
\Lambda_{2n} \cong \bigotimes_{j=1}^{n}
\Lambda(\bbC\grx_{2j-1}\oplus\bbC\grx_{2j}) \mcomma
\]
we have
\[
\Uf^{\sigma} = G_{\sigma}^{\otimes n} \mcomma \qquad
(\Uf^{\sigma})^2 = P_f \mcomma \qquad
(\Uf^{\sigma})^4 = \Id \mdot
\]
In particular, $\Uf^{\sigma}$ is unitary with respect to the occupation number inner product, and satisfies $(\Uf^{\sigma})^\dagger = (\Uf^{\sigma})^{-1} = \Uf^{-\sigma}$. (In contrast, the unscaled transform $\Gf^\sigma$ maps $1$ to $2^{-n}\gry_1\cdots\gry_{2n}$, which has norm $2^{-n} \neq 1$ for $n\ge1$.)

Since $\Dn\partial_{\grx_a}\Dn^{-1} = 2^{1/2}\partial_{\grx_a}$ and $\Dn^{-1}\grx_a\Dn = 2^{1/2}\grx_a$, conjugating the intertwining relations in Proposition \ref{prop:intertwining} by $\Dn$ removes the factors of $2$ and $\tfrac12$ and gives the symmetric intertwining relations:
\begin{align}
\Uf^{\sigma}\partial_{\grx_{2j}}
&=-\sigma\compi\,\gry_{2j-1}\Uf^{\sigma}\mcomma
&
\Uf^{\sigma}\partial_{\grx_{2j-1}}
&=\sigma\compi\,\gry_{2j}\Uf^{\sigma}\mcomma\label{eq:Uf-ann}\\
\Uf^{\sigma}\grx_{2j}
&=\sigma\compi\,\partial_{\gry_{2j-1}}\Uf^{\sigma}\mcomma
&
\Uf^{\sigma}\grx_{2j-1}
&=-\sigma\compi\,\partial_{\gry_{2j}}\Uf^{\sigma}\mdot\label{eq:Uf-cre}
\end{align}
Since $\Dn$ commutes with $P_f$, we also have
\[
P_f\Uf^{\sigma} = \Uf^{-\sigma} \mcomma \qquad
\Gf^{\sigma} = \Dn\Uf^{\sigma}\Dn^{-1} \mdot
\]
Combining $\Ub^\sigma$ and $\Uf^\sigma$, we define the unitarily normalized hybrid Fourier transform by
\begin{equation}\label{eq:hybrid-gate-factor}
\Uhyb^{\sigma} \coloneqq \Ub^{\sigma}\otimes\Uf^{\sigma} \mdot
\end{equation}
Then $\Uhyb^\sigma$ is a unitary operator on $\caH_{m|2n}$ that preserves $\caS(\bbR^{m|2n})$. Writing $W \coloneqq \Id_{\caS(\bbR^m)}\otimes\Dn$, we have the similarity relation
\begin{equation}\label{eq:raw-unitary-similarity}
\Fhyb^{\sigma} = \Ub^{\sigma}\otimes\Gf^{\sigma}
= W\,\Uhyb^{\sigma}\,W^{-1} \mdot
\end{equation}

For $\sigma=+1$, let us denote by $\kappa_F \coloneqq \operatorname{Conj}(\Uhyb^+)$ the conjugation action of $\Uhyb^+$. By (\ref{eq:Uf-ann}) and (\ref{eq:Uf-cre}), $\kappa_F$ acts on the first-order canonical generators by
\begin{align*}
\kappa_F(x_r)&=-\compi\,\partial_{y_r}\mcomma &
\kappa_F(\partial_{x_r})&=-\compi\,y_r\mcomma\\
\kappa_F(\grx_{2j})&=\compi\,\partial_{\gry_{2j-1}}\mcomma &
\kappa_F(\grx_{2j-1})&=-\compi\,\partial_{\gry_{2j}}\mcomma\\
\kappa_F(\partial_{\grx_{2j}})&=-\compi\,\gry_{2j-1}\mcomma &
\kappa_F(\partial_{\grx_{2j-1}})&=\compi\,\gry_{2j}\mdot
\end{align*}
More generally, for any differential operator $Z$ in the Weyl--Clifford algebra generated by $\{x_r,\partial_{x_r},\grx_a,\partial_{\grx_a}\}$, if we write $W_x$ and $W_y$ for the scaling operator $W$ acting in the source and target variables, we have
\[
\kappa_F(Z)
= W_y^{-1}\kappa(W_xZW_x^{-1})W_y \mcomma
\qquad
\Uhyb^+(Zf) = \kappa_F(Z)(\Uhyb^+f)
\]
for all $f\in\caS(\bbR^{m|2n})$, where $\kappa=\operatorname{Conj}(\Fhyb^+)$ is the conjugation map from Section 4.4.

\subsection{Rigidity of the normalized transform}\label{subsec:unitary-intertwiners}

In this subsection, we again identify the source and target variables so that $W_x = W_y = W$. By transporting Theorem \ref{thm:generalFT} via $W$, we obtain the following unitary rigidity result for $\Uhyb^+$.

\begin{prop}\label{corl:unitary-family}
Assume $m,n\ge1$, and let $T$ be a continuous complex-linear operator on $\caS(\bbR^{m|2n})$ satisfying the coordinate quadratic intertwining relations
\[
TQ = \kappa_F(Q)T \qquad (Q\in\mathcal Q) \mcomma
\]
where $\mathcal Q$ is the quadratic family in (\ref{eq:quadratic-family}) (or equivalently, full even covariance $TQ = \kappa_F(Q)T$ for all $Q\in\frg_0$ together with $TZ=\kappa_F(Z)T$ for one fixed nonzero $Z\in\frg_1$).
Then there exist unique scalars $\alpha,\beta\in\bbC$ such that
\[
T = \alpha\,\Uhyb^{+}+\beta\,\Uhyb^{-}
= (\alpha\,\Id+\beta\,\Pbf)\Uhyb^{+} \mdot
\]
Moreover, $T$ extends to a unitary operator on $\caH_{m|2n}$ if and only if
\[
|\alpha+\beta| = |\alpha-\beta| = 1 \mcomma
\]
i.e., if and only if there exist real numbers $\phi,\vartheta\in\bbR$ such that
\[
T = e^{\compi\phi}e^{\compi\vartheta\Pbf}\Uhyb^{+} \mdot
\]
\end{prop}

\begin{proof}
Since conjugation by $W = \Id\otimes\Dn$ rescales each coordinate generator in $\mathcal Q$ by a nonzero constant and preserves the subspaces $\frg_0$ and $\frg_1$, the relation $TQ = \kappa_F(Q)T$ is equivalent to
\[
(WTW^{-1})Q = \kappa(Q)(WTW^{-1}) \mdot
\]
By Theorem \ref{thm:generalFT}(2), we obtain $WTW^{-1} = \alpha\Fhyb^{+}+\beta\Fhyb^{-}$ for unique scalars $\alpha,\beta\in\bbC$. Using (\ref{eq:raw-unitary-similarity}) and the fact that $W$ commutes with $\Pbf$, we get
\[
T = \alpha\Uhyb^+ + \beta\Uhyb^- = (\alpha\Id + \beta\Pbf)\Uhyb^+ \mdot
\]
Since $\Uhyb^{+}$ is unitary on $\caH_{m|2n}$ and $\Pbf$ has eigenvalues $+1$ and $-1$ on the two nonzero total-parity sectors, $T$ is unitary if and only if $|\alpha+\beta|=|\alpha-\beta|=1$. Writing $\alpha+\beta = e^{\compi(\phi+\vartheta)}$ and $\alpha-\beta = e^{\compi(\phi-\vartheta)}$ for $\phi,\vartheta\in\bbR$, we get $\alpha\Id+\beta\Pbf = e^{\compi\phi}e^{\compi\vartheta\Pbf}$. This finishes the proof.
\end{proof}

\begin{corl}\label{corl:one-odd-test}
Under the assumptions of Proposition \ref{corl:unitary-family}, suppose in addition that $T$ is unitary and satisfies any one nonzero first-order intertwining relation
\[
TZ = \kappa_F(Z)T
\]
for a canonical generator $Z \in E_b \oplus E_f$. Then we have
\[
T = e^{\compi\phi}\Uhyb^{+}
\]
for some $\phi\in\bbR$.
\end{corl}

\begin{proof}
By Proposition \ref{corl:unitary-family}, we can write $T=e^{\compi\phi}e^{\compi\vartheta\Pbf}\Uhyb^{+}$. Since $\Uhyb^{+}Z=\kappa_F(Z)\Uhyb^{+}$, the condition $TZ=\kappa_F(Z)T$ is equivalent to $[e^{\compi\vartheta\Pbf},\, \kappa_F(Z)] = 0$. Since every first-order canonical generator in $E_b\oplus E_f$ reverses total parity, $\kappa_F(Z)$ anticommutes with $\Pbf$:
\[
\Pbf\,\kappa_F(Z) = -\kappa_F(Z)\Pbf \mdot
\]
Therefore, we have
\[
[e^{\compi\vartheta\Pbf},\, \kappa_F(Z)]
= 2\compi\sin(\vartheta)\,\Pbf\,\kappa_F(Z) \mdot
\]
Since $\kappa_F(Z)\neq 0$, this commutator vanishes if and only if $\sin(\vartheta)=0$, which means $e^{\compi\vartheta\Pbf}=\pm\Id$. Absorbing the sign into $e^{\compi\phi}$, we get our result.
\end{proof}

\begin{remark}
Let $m,n\ge1$, put $T_\vartheta \coloneqq e^{\compi\vartheta\Pbf}\Uhyb^+$, and write $q_r \coloneqq x_r$ and $p_r \coloneqq -\compi\partial_{x_r}$. Since $\Uhyb^+q_r(\Uhyb^+)^\dagger = p_r$ and $\Pbf p_r = -p_r\Pbf$, expanding $e^{\compi\vartheta\Pbf}=\cos\vartheta\,\Id+\compi\sin\vartheta\,\Pbf$ gives
\begin{equation}\label{eq:parity-dressed-quadrature}
T_\vartheta q_rT_\vartheta^\dagger
= \cos(2\vartheta)p_r
+\compi\sin(2\vartheta)\Pbf p_r \mdot
\end{equation}
Note that $\Pbf p_r$ does not lie in the first-order space $E_b\oplus E_f$: it commutes with $P_f$, yet acts as $+P_bp_r$ on the even fermionic sector and as $-P_bp_r$ on the odd fermionic sector. Thus, whenever $\sin(2\vartheta)\ne0$, conjugation by $T_\vartheta$ does not preserve $E_b\oplus E_f$, even though $T_\vartheta$ satisfies all quadratic intertwining relations.
\end{remark}

In physical applications, we distinguish the total parity $\Pbf=P_bP_f$ from the fermion parity $P_f$. Under fermion-parity superselection, physical observables must commute with $P_f$ (cf. \cite{Friis2013}), whereas a mixed quadratic operator $x_r\grx_a$ anticommutes with $P_f$. Note that in Corollary \ref{corl:one-odd-test}, we can choose the single first-order test operator to be $Z=q_r$, which commutes with $P_f$.

\subsection{A rotation in the Lefschetz representation}\label{subsec:lefschetz-realization}

On the occupation number Hilbert space $\Lambda_{2n}$, let us write
\[
c_a \coloneqq \partial_{\grx_a} \mcomma \qquad c_a^\dagger \coloneqq \grx_a \mcomma \qquad
n_a \coloneqq c_a^\dagger c_a \mcomma \qquad N_f \coloneqq \sum_{a=1}^{2n}n_a \mdot
\]
With respect to the orthonormal monomial basis $\{\grx_I\}$, $c_a^\dagger$ is the Hilbert space adjoint of $c_a$. For each symplectic pair $1\le j\le n$, we define the self-adjoint pairing operator
\begin{equation}\label{eq:pairing-generator}
K_j \coloneqq c_{2j-1}^\dagger c_{2j}^\dagger+c_{2j}c_{2j-1} \mdot
\end{equation}
On the $j$-th pair, $K_j$ exchanges $|00\rangle$ and $|11\rangle$ and annihilates $|01\rangle$ and $|10\rangle$, and $K_1,\ldots,K_n$ commute with one another.

Recall the Lefschetz operators $X,Y,H$ from Lemma \ref{lem:dual_pair}. Conjugating them by $\Dn$, we define
\[
J_+ \coloneqq \sum_{j=1}^n c_{2j-1}^\dagger c_{2j}^\dagger
= 2X = \Dn^{-1}X\Dn \mcomma \qquad
J_- \coloneqq J_+^\dagger = Y/2 = \Dn^{-1}Y\Dn \mcomma
\]
\[
J_z \coloneqq (N_f-n)/2 = H/2 \mcomma \qquad
J_x \coloneqq (J_++J_-)/2 \mcomma \qquad J_y \coloneqq (J_+-J_-)/(2\compi) \mdot
\]
By the $\fsl_2$-relations for $X,Y,H$, we have $[J_+,J_-]=2J_z$ and $[J_z,J_\pm]=\pm J_\pm$. Since $J_+^\dagger=J_-$ and $J_z^\dagger=J_z$, the operators $J_x,J_y,J_z$ are self-adjoint on $\Lambda_{2n}$, and $\compi J_x,\compi J_y,\compi J_z$ span a unitary representation of $\fsu(2)$ (which integrates to $SU(2)$). Conjugating De Bie's exponential formula \cite[Theorem 10]{DB08} (see the proof of Proposition \ref{prop:fourier_decomp}) by $\Dn$, we obtain
\begin{equation}\label{eq:pairing-pulse}
\Uf^\sigma = (\sigma\compi)^n
\exp\!\left(-\frac{\sigma\compi\pi}{2}\sum_{j=1}^nK_j\right)
= (\sigma\compi)^n e^{-\sigma\compi\pi J_x} \mdot
\end{equation}
We can also check (\ref{eq:pairing-pulse}) directly from Proposition \ref{prop:onepair}: on the even subspace of the $j$-th pair, $K_j$ has matrix $\left(\begin{smallmatrix}0&1\\1&0\end{smallmatrix}\right)$, while $K_j$ vanishes on the odd subspace; hence $(\sigma\compi)e^{-\sigma\compi\pi K_j/2} = G_\sigma$, and taking the tensor product over $1\le j\le n$ gives (\ref{eq:pairing-pulse}). Thus, up to the scalar phase $(\sigma\compi)^n$, $\Uf^\sigma$ is an $SU(2)$ rotation in the Lefschetz representation.

On each irreducible summand $\Dn^{-1}(V_{n-k}\otimes\caH_k^f)$, this $SU(2)$ rotation acts on the first factor in the spin-$\frac{n-k}{2}$ representation and acts as the identity on the harmonic factor $\caH_k^f$. Moreover, for a single symplectic pair in the ordered occupation basis, the even block of $G_\sigma$ is $A=\left(\begin{smallmatrix}0&1\\1&0\end{smallmatrix}\right)$ and the odd block is $B=\sigma\compi I_2$. Since $\det A=\det B=-1$, $G_\sigma$ is a standard two-qubit matchgate in the sense of \cite{JozsaMiyake2008}. Note that $\Uf^\sigma$ preserves the fermion parity $P_f$, and sends the total fermionic number operator $N_f$ to $\Uf^\sigma N_f(\Uf^\sigma)^\dagger = 2n\Id-N_f$.

\subsection{The normalized transform on hybrid Hamiltonians}
\label{subsec:density-quadrature}
Now we apply the normalized hybrid Fourier transform $\Uhyb^\sigma$ to Hamiltonians of hybrid $2n$-qubit--$m$-qumode quantum systems. Let us put $q_r \coloneqq x_r$, $p_r \coloneqq -\compi\partial_{x_r}$, and $N_b \coloneqq \frac12\sum_{r=1}^m(q_r^2+p_r^2-\Id)$. From the Hermite eigenvalues of $\Ub^\sigma$ and (\ref{eq:pairing-pulse}), we have
\[
 \Ub^\sigma = e^{\sigma\compi\pi N_b/2} \mcomma \qquad
 \Uhyb^\sigma
 = (\sigma\compi)^n e^{\sigma\compi\pi N_b/2}
 e^{-\sigma\compi\pi\sum_{j=1}^n K_j/2} \mdot
\]
By Proposition \ref{prop:bosonicFourier1} and (\ref{eq:Uf-ann})--(\ref{eq:Uf-cre}), conjugation by $\Uhyb^\sigma$ acts on the quadrature operators and fermionic occupation numbers by
\begin{align}
 \Uhyb^\sigma q_r(\Uhyb^\sigma)^\dagger &= \sigma p_r \mcomma
 &\Uhyb^\sigma p_r(\Uhyb^\sigma)^\dagger &= -\sigma q_r \mcomma
 \label{eq:physical-quadrature-action}\\
 \Uf^\sigma n_{2j-1}(\Uf^\sigma)^\dagger &= \Id-n_{2j} \mcomma
 &\Uf^\sigma n_{2j}(\Uf^\sigma)^\dagger &= \Id-n_{2j-1} \mdot
 \label{eq:physical-number-action}
\end{align}

Let us consider the case $m=1$ and $n\ge1$. Write $q=q_1$, $p=p_1$, and $Q_f \coloneqq N_f-n\Id$. For $\omega>0$ and $\mu,g\in\bbR$, consider the density--displacement Hamiltonian
\begin{equation}
 H \coloneqq \frac{\omega}{2}(q^2+p^2)+\mu Q_f+gqQ_f \mdot
 \label{eq:density-displacement-hamiltonian}
\end{equation}
Note that the interaction term $gqQ_f$ preserves the fermion parity $P_f$. On each eigenspace of $Q_f$ with eigenvalue $\lambda$, we can complete the square to write
\[
 H = \frac{\omega}{2}
 \left[p^2+\left(q+\frac{g\lambda}{\omega}\right)^2\right]
 +\mu\lambda-\frac{g^2\lambda^2}{2\omega} \mcomma
\]
which is a shifted harmonic oscillator plus a constant and is essentially self-adjoint on $\caS(\bbR)$. Since $\Lambda_{2n}$ is finite-dimensional, $H$ is essentially self-adjoint on $\caS(\bbR)\otimes\Lambda_{2n}$. By (\ref{eq:physical-number-action}), we have $\Uf^\sigma Q_f(\Uf^\sigma)^\dagger = -Q_f$, and hence conjugating $H$ by $U_{1|2n}^\sigma$ gives the density--momentum Hamiltonian:
\begin{equation}
 U_{1|2n}^\sigma H(U_{1|2n}^\sigma)^\dagger
 = \frac{\omega}{2}(q^2+p^2)-\mu Q_f-\sigma g pQ_f \mdot
 \label{eq:density-momentum-duality}
\end{equation}
In particular, the two Hamiltonians in (\ref{eq:density-displacement-hamiltonian}) and (\ref{eq:density-momentum-duality}) are unitarily equivalent and have the exact same spectrum.

\begin{prop}
\label{corl:physical-number-identification}
Assume $m=1$ and $n\ge2$. Let $T$ be a unitary operator on $L^2(\bbR)\widehat\otimes\Lambda_{2n}$ that preserves $\caS(\bbR)\otimes\Lambda_{2n}$ and acts continuously on $\caS(\bbR)\otimes\Lambda_{2n}$. Suppose that
\[
 TQ = \kappa_F(Q)T \quad (Q\in\frg_0) \mcomma \qquad
 T(qQ_f) = -pQ_fT \mcomma \qquad \kappa_F = \operatorname{Conj}(U_{1|2n}^+) \mdot
\]
Then we have $T = e^{\compi\phi}e^{\compi\theta P_f}U_{1|2n}^+$ for some $\phi,\theta\in\bbR$.
\end{prop}
\begin{proof}
Put $A \coloneqq (U_{1|2n}^+)^{-1}T$. Since $T$ satisfies full even covariance $TQ = \kappa_F(Q)T$ for all $Q\in\frg_0$, it follows from Lemma \ref{lem:quadratic-oscillator} and similarity by $W=\Id\otimes\Dn$ that $A = \sum_{s,t\in\{\pm1\}} a_{s,t}\Pi_{s,t}$. Moreover, since $U_{1|2n}^+(qQ_f)(U_{1|2n}^+)^{-1} = -pQ_f$, the condition $T(qQ_f) = -pQ_fT$ is equivalent to $[A,\, qQ_f] = 0$.
When $n\ge2$, although $Q_f = N_f-n\Id$ vanishes on the degree-$n$ subspace, it is nonzero on both the even and odd fermionic parity sectors: indeed, on the vacuum $1 \in \caF_+$ it takes the value $-n \neq 0$, and on a degree-$1$ monomial $\grx_1 \in \caF_-$ it takes the value $1-n \neq 0$. Meanwhile, multiplication by $q$ flips the bosonic parity $P_b$ and is injective on $\caS(\bbR)$. Therefore, $[A,\, qQ_f] = 0$ forces $a_{+,+} = a_{-,+}$ on the even fermionic sector and $a_{+,-} = a_{-,-}$ on the odd fermionic sector. This means that $A = a\Id + bP_f$ for some $a,b\in\bbC$, so $T = (a\Id+bP_f)U_{1|2n}^+$. Finally, since $T$ and $U_{1|2n}^+$ are unitary, we have $|a+b|=|a-b|=1$, which gives $T = e^{\compi\phi}e^{\compi\theta P_f}U_{1|2n}^+$ for some $\phi,\theta\in\bbR$. This finishes the proof.
\end{proof}

\begin{remark}
For a standalone physical system subject to fermion-parity superselection (cf. \cite{Friis2013, Szalay2021, Vidal2021}), any physical density operator $\rho$ and any bounded observable $O$ commute with $P_f$. Thus $T = e^{\compi\phi}e^{\compi\theta P_f}\Uhyb^+$ and $\Uhyb^+$ induce the exact same conjugation action:
\[
 T\rho T^\dagger = \Uhyb^+\rho(\Uhyb^+)^\dagger \mcomma \qquad
 T^\dagger OT = (\Uhyb^+)^\dagger O\Uhyb^+ \mdot
\]
Note also that when $n=1$, $Q_f = N_f - \Id$ vanishes identically on the odd fermionic sector $\caF_- = \vectorspan\{\grx_1,\grx_2\}$, so Proposition \ref{corl:physical-number-identification} requires $n\ge2$, whereas the Hamiltonian duality (\ref{eq:density-momentum-duality}) holds for all $n\ge1$.
\end{remark}

\begin{exmp}
Let $m=1$, $n\ge1$, and $Q_f=N_f-n\Id$. For any $\tau\in\bbR$, by applying the spectral functional calculus to (\ref{eq:physical-quadrature-action}) and (\ref{eq:physical-number-action}), we obtain the unitary conjugation formula
\begin{equation}\label{eq:conditional-displacement-conjugation}
 U_{1|2n}^\sigma e^{-\compi \tau qQ_f}(U_{1|2n}^\sigma)^\dagger
 = e^{\sigma\compi \tau pQ_f} \mdot
\end{equation}
When $n=1$, if we restrict the fermionic factor to the even-parity subspace $\vectorspan\{|00\rangle,|11\rangle\}$, the operator $e^{-\compi \tau qQ_f}$ acts as a conditional displacement gate:
\[
 e^{-\compi \tau qQ_f}
 = e^{\compi \tau q}\otimes|00\rangle\langle00|
  +e^{-\compi \tau q}\otimes|11\rangle\langle11| \mdot
\]
Such conditional displacements are basic building blocks in hybrid oscillator--qubit simulation (cf. \cite[Section III.3.1]{Crane2026}), and (\ref{eq:conditional-displacement-conjugation}) exchanges the two quadrature directions and reverses the controlling centered density.
\end{exmp}

\subsection{A final remark on Hubbard--Holstein models}

Finally, let us consider the Hubbard--Holstein model on $n\ge1$ sites, with two spin states of a fermion and one bosonic mode at each site. Its occupation number encoding uses $2n$ qubits and $n$ qumodes, and its common invariant domain is $\caS(\bbR^n)\otimes\Lambda_{2n}$, so $m=n$. For each site $1\le j\le n$, we denote the local centered fermionic density by $\rho_j \coloneqq n_{2j-1}+n_{2j}-\Id$. For real parameters $t,u,\mu,g\in\bbR$ and $\omega>0$, consider the open-chain Hubbard--Holstein Hamiltonian
\begin{align}
 H \coloneqq {}&-t\sum_{a=1}^{2n-2}
       (c_a^\dagger c_{a+2}+c_{a+2}^\dagger c_a)
  +u\sum_{j=1}^{n}
       (n_{2j-1}-\tfrac12\Id)(n_{2j}-\tfrac12\Id)\notag\\
 &+\frac\omega2\sum_{j=1}^{n}(q_j^2+p_j^2)
       -\mu\sum_{j=1}^{n}\rho_j
       +g\sum_{j=1}^{n} q_j\rho_j \mdot
 \label{eq:hubbard-holstein-chain}
\end{align}
These electron--phonon models are important targets for hybrid oscillator--qubit quantum simulation (cf. \cite{Chen2025Genesis}).
By (\ref{eq:Uf-ann}) and (\ref{eq:Uf-cre}), we have
\[
 \Uf^\sigma c_{2j-1}(\Uf^\sigma)^\dagger
 = \sigma\compi c_{2j}^\dagger \mcomma \qquad
 \Uf^\sigma c_{2j}(\Uf^\sigma)^\dagger
 = -\sigma\compi c_{2j-1}^\dagger \mdot
\]
Combining these relations with the canonical anticommutation relations and (\ref{eq:physical-quadrature-action}), we obtain the conjugated Hamiltonian
\begin{align}
 U_{n|2n}^\sigma H(U_{n|2n}^\sigma)^\dagger
 ={}&t\sum_{a=1}^{2n-2}
       (c_a^\dagger c_{a+2}+c_{a+2}^\dagger c_a)
  +u\sum_{j=1}^{n}
       (n_{2j-1}-\tfrac12\Id)(n_{2j}-\tfrac12\Id)\notag\\
 &+\frac\omega2\sum_{j=1}^{n}(q_j^2+p_j^2)
       +\mu\sum_{j=1}^{n}\rho_j
       -\sigma g\sum_{j=1}^{n} p_j\rho_j \mdot
 \label{eq:hubbard-holstein-conjugation}
\end{align}

\begin{remark}
The space $\caS(\bbR^n)\otimes\Lambda_{2n}$ carries the quadratic oscillator action of $\fosp(4n|2n)$. For $n\ge2$, Nishiyama \cite[Section 4]{Nis91} studies a different oscillator module for a real form $\fosp(2n/4n;\bbR)$  (in \cite{Nis91} the symplectic dimension was placed first) of $\fosp(4n|2n)$ and its restriction to a real super dual pair whose complexification is $(\fosp(1|2),\fso(2n))$. It remains to establish the connection of 
this new dual pair decomposition to the hybrid oscillator--qubit quantum simulation in the Hubbard--Holstein models, which we plan to do elsewhere.
\end{remark}

\section*{Acknowledgements}
J.-S. Huang is supported by NSFC grants 11271460 \& 12341101, Guangdong provincial grants 2024A1515015146 \& GDZX2403006, and Shenzhen municipal grant JCYJ20241202124023031.
Y. Luan is supported by the Shandong Provincial Natural Science Foundation (Grant No. ZR2025MS36).

\section*{Declaration of using AI}
During the preparation of this work, the authors used AI for language polishing and exploratory brainstorming for possible approaches. AI-generated suggestions were not treated as mathematical evidence. The authors independently constructed and verified every mathematical argument, checked every citation, reviewed and edited the resulting text, and take full responsibility for the content of the article.

\bibliography{HybridFourierTransform}
\bibliographystyle{amsplain}
\end{document}